\documentclass[onecolumn,amsthm]{autart_arxiv}

\newif\ifFastBW
\FastBWtrue

\newif\ifTabSideBySide
\TabSideBySidetrue

\usepackage{graphicx}
\usepackage{enumitem}
\usepackage{nth}
\usepackage{balance}
\usepackage{subcaption}
\usepackage{placeins}
\usepackage[binary-units]{siunitx}

\usepackage[sort&compress,numbers]{natbib}

\usepackage{xcolor}
\definecolor{cblue}{HTML}{045275}
\definecolor{cred}{HTML}{F0746E}
\definecolor{cgreen}{HTML}{7CCBA2}
\definecolor{legreddraw}{HTML}{E41A1C}
\definecolor{legbluedraw}{HTML}{377EB8}
\definecolor{leggreendraw}{HTML}{4DAF4A}
\definecolor{legvioletdraw}{HTML}{984EA3}
\definecolor{legredfill}{HTML}{B61516}
\definecolor{legbluefill}{HTML}{2C6593}
\definecolor{leggreenfill}{HTML}{3E8C3B}
\definecolor{legvioletfill}{HTML}{7A3E82}

\usepackage{amsmath, amssymb, amsfonts, mathtools, bm, amsthm}
\usepackage{stmaryrd} 
\usepackage{mleftright}
\usepackage{mymath} 
\usepackage{arydshln} 

\usepackage{etoolbox}
\usepackage{algorithm}
\let\classAND\AND
\let\AND\relax
\usepackage{algorithmic}

\let\AND\classAND
\AtBeginEnvironment{algorithmic}{\let\AND\algoAND}
\usepackage{algorithmic}

\usepackage{parskip} 
\begingroup
    \makeatletter
    \@for\theoremstyle:=definition,remark,plain\do{%
        \expandafter\g@addto@macro\csname th@\theoremstyle\endcsname{%
            \addtolength\thm@preskip\parskip
            }%
        }
\endgroup

\usepackage{tabularx,ragged2e,booktabs,caption,multirow} 
\usepackage{array}
\usepackage{makecell} 

\usepackage{extarrows}
\usepackage{enumerate}
\usepackage{enumitem}
\RequirePackage{rotating} 
\RequirePackage{fix-cm}
\usepackage{stackengine}
\usepackage{multicol}
\usepackage{rotating}
\usepackage{booktabs}

\usepackage{pgfplots}
\usepgfplotslibrary{patchplots,colormaps}
\pgfplotsset{compat=newest}
\pgfplotsset{
  layers/axis lines on top/.define layer set={
    axis background,
    axis grid,
    axis ticks,
    axis tick labels,
    pre main,
    main,
    axis lines,
    axis descriptions,
    axis foreground,
  }{/pgfplots/layers/standard},
}

\usepgfplotslibrary{colorbrewer}

\usepackage{tikz}
\usetikzlibrary{shapes, arrows, arrows.meta, positioning, calc,pgfplots.colorbrewer}
\newlength{\Llong}
\newlength{\Lmid}
\newlength{\Lshort}
\usepackage{csvsimple}
\usetikzlibrary{external}
\newcommand{\inputtikz}[1]{%
\centering
\includegraphics[scale=1]{figures/#1.pdf}
}

\usetikzlibrary{plotmarks}
\newenvironment{customlegend}[1][]{%
    \begingroup
    \csname pgfplots@init@cleared@structures\endcsname
    \pgfplotsset{#1}%
}{%
    \csname pgfplots@createlegend\endcsname
    \endgroup
}%
\def\addlegendimage{\csname pgfplots@addlegendimage\endcsname}
\newcommand{\addlegendimageintext}[1]{%
    \tikzexternaldisable
    \tikz {
        \begin{customlegend}[
            legend entries={\empty},
            legend style={
                at={(current bounding box.north west)},
    			anchor=north west,
                draw=none,
                inner sep=1.5pt,
                column sep=0pt,
                nodes={inner sep=0pt}}]
        \addlegendimage{#1}
        \end{customlegend}
    }%
    \tikzexternalenable
}

\usepackage[normalem]{ulem}

\begin{document}

\begin{frontmatter}

\title{Control of Cross-Directional Systems\\ with Approximate Symmetries\thanksref{footnoteinfo}}

\thanks[footnoteinfo]{Corresponding author I. Kempf.}

\author[I]{Idris Kempf}\ead{idris.kempf@eng.ox.ac.uk}, 
\author[I]{Paul Goulart}\ead{paul.goulart@eng.ox.ac.uk},
\author[I]{Stephen Duncan}\ead{stephen.duncan@eng.ox.ac.uk}

\address[I]{Department of Engineering Science, University of Oxford, Oxford OX1 3PJ, UK}

\begin{keyword}
Spatially invariant systems, cross-directional systems, synchrotrons, pulp and paper industry
\end{keyword}

\begin{abstract}
Structural symmetries of linear dynamical systems can be exploited for decoupling the dynamics and reducing the computational complexity of the controller implementation. However, in practical applications, inexact structural symmetries undermine the ability to decouple the system, resulting in the loss of any potential complexity reduction. To address this, we propose substituting an approximation with exact structural symmetries for the original system model, thereby introducing an approximation error. We focus on internal model controllers for cross-directional systems encountered in large-scale and high-speed control problems of synchrotrons or the process industry and characterise the stability, performance, and robustness properties of the resulting closed loop. While existing approaches replace the original system model with one that minimises the Frobenius norm of the approximation error, we show that this can lead to instability or poor performance. Instead, we propose approximations that are obtained from semidefinite programming problems. We show that our proposed approximations can yield stable systems even when the Frobenius norm approximation does not. The paper concludes with numerical examples and a case study of a synchrotron light source with inexact structural symmetries. Exploiting structural symmetries in large-scale and high-speed systems enables faster sampling times and the use of more advanced control techniques, even when the symmetries are approximate.
\end{abstract}

\end{frontmatter}

\section{Introduction\label{sec:introduction}}
Symmetry is always accompanied by redundancies in the mathematical representation of dynamical systems and can be exploited to reduce the complexity of the system model~\cite{MAIDENS2018367}. Many real-world systems naturally exhibit symmetries, which has been taken advantage of in control problems ranging from process engineering~\cite{PAPERMACHINES} to distributed control~\cite{RAHMANI}. A set of matrices with a \emph{block-structural symmetry} is defined as the subspace
\begin{align}\label{eq:sym}
\mathcal{S}\eqdef\left\lbrace
\inC{R}{n{b_y}}{n{b_u}}
\mid
R\left(\Pi\otimes I_{b_u}\right) = \left(\Pi\otimes I_{b_y}\right) R
\right\rbrace,
\end{align}
which is associated with a \emph{permutation matrix} $\inR{\Pi}{n}{n}$~\cite[Ch.\ 5]{GTBISHOP}. All $R\in\mathcal{S}$ are block-diagonalised by the same orthonormal $\inC{V}{n}{n}$ and form a commutative algebra (Section~\ref{sec:background}). Structural symmetry can be extended to linear dynamical systems~\cite{APPROXSYMMPC}, in which case most controllers inherit the symmetry properties. When $V$ is a computationally efficient transformation, such as the Fast Fourier Transformation (FFT) for \emph{circulant} matrices~\cite{CIRCBOOK}, structural symmetries can also be exploited to speed up controller computations, which is particularly useful for large-scale and high-speed systems~\cite{ANDREADISTRIBUTED,BCINVADMMCONF,PHDDANIELSON}.

Structural symmetries are frequently encountered in \emph{cross-directional} (CD) systems~\cite{HEATHWILLS},
\begin{align}\label{eq:CDsystem}
y(s) = R g(s) u(s) + d(s),
\end{align}
where $s\in\C$ is the Laplace variable, $g:\C\rightarrow\C$ the scalar actuator dynamics, $u:\C\rightarrow\C^{n{b_u}}$ are the control inputs, $y:\C\rightarrow\C^{n{b_y}}$ the outputs and $d:\C\rightarrow\C^{n{b_y}}$ the disturbances. If the sensors and actuators of a CD system are arranged in a regular pattern, then the \emph{response matrix} $\inC{R}{n{b_y}}{n{b_u}}$ inherits a (possibly partial) structural symmetry~\cite{CROSSDIR}. However, most systems in practice only adhere approximately to a structural symmetry~\cite{SAJJADROBUST,APPROXSYMMPC,SYNCSYM}, meaning that
\begin{align}\label{eq:approxsymmetry}
\norm{R\left(\Pi\otimes I_{b_u}\right)-\left(\Pi\otimes I_{b_y}\right) R}\reqdef\epsilon >0,
\end{align}
where $\anynorm{\cdot}$ is an arbitrary norm. In this case, $\hat{R}\eqdef(\herm{V}\otimes I_{b_y})R(V\otimes I_{b_u})$ is \textit{not} block-diagonal, so that the advantages of the transformation into symmetric domain are lost. To recover the structural symmetry of~\eqref{eq:CDsystem} and the associated speed advantages, one possibility is to split $R$ as
\begin{align}\label{eq:Rsdef}
R = R_\mathcal{S}+\Delta_\mathcal{S},
\end{align}
with $R_\mathcal{S}\in\mathcal{S}$, thereby artificially introducing an approximation error $\Delta_\mathcal{S}\eqdef R-R_\mathcal{S}$. A robust controller $Q:\C\mapsto\C^{n b_u \times n b_y}$ can then be designed using $P_\mathcal{S}(s)\eqdef R_\mathcal{S} g(s)$ and used to control the real plant $P(s)\eqdef Rg(s)$. Another possibility is to enforce the constraint $Q(s)\in\mathcal{S}$ $\forall s\in\C$ during synthesis, which is analogous to the design of decentralised controllers~\cite[Ch.\ 12]{morari1989robust}. However, the constraint $Q(s)\in\mathcal{S}$ $\forall s$ can lead to a non-convex optimisation problem~\cite{QUADINV,GROEBNER1} and this method is not further considered.

\begin{figure}
\inputtikz{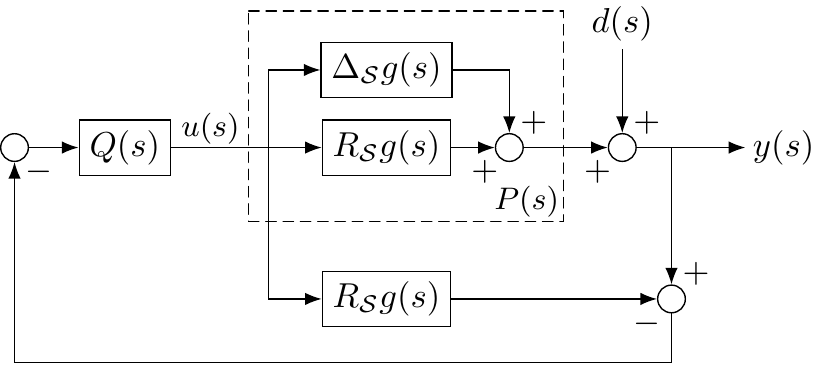}
\caption{IMC structure with the approximation $R_\mathcal{S}$ and approximation error $\Delta_\mathcal{S}$, where $P(s)=(R_\mathcal{S}+\Delta_\mathcal{S})g(s)$ is known and $\Delta_\mathcal{S}$ artificially introduced through $R_\mathcal{S}$.\label{fig:fbsystem}}
\end{figure}

In this paper, it is assumed that $R$ satisfies~\eqref{eq:approxsymmetry} for some $\epsilon>0$ and the feedback structure is fixed to the \emph{internal model control} (IMC) structure~\cite[Ch.\ 3]{morari1989robust} from Fig.~\ref{fig:fbsystem}. The IMC filter $Q(s)$ is assumed to be a \emph{Dahlin} or \emph{lambda} controller~\cite[Ch.\ 4.5]{morari1989robust},
\begin{align}\label{eq:Qimc}
&Q(s)\eqdef\pinv{R}q(s),
&q(s)\eqdef T(s) / g(s),
\end{align}
where $\pinv{R}$ is the Moore-Penrose pseudoinverse of $R$~\cite[P5.5.2]{GOLUB4} and $T:\C\mapsto\C$ the \emph{complementary sensitivity}, which must include the non-minimum phase parts of $g(s)$. 

With the controller structure fixed, the first aim of this paper is to analyse~\eqref{eq:CDsystem} when an approximation $R_\mathcal{S}$~\eqref{eq:Rsdef} is substituted for $R$ in~\eqref{eq:Qimc}, i.e.\ when the IMC filter is re-defined as
\begin{align}\label{eq:Q}
Q(s)\eqdef \pinv{R_\mathcal{S}} q(s),
\end{align}
and embedded in the IMC structure from Fig.~\ref{fig:fbsystem}. By choosing the controller as in~\eqref{eq:Q}, the commutative algebra of the matrices in $\mathcal{S}$ allows the structure of only $R$ to be constrained, which is then inherited by $Q(s)$.

One approximation that has been used in this setting is $R_\mathcal{S}^\text{F} \eqdef \argmin_{X\in\,\mathcal{S}} \frobnorm{X- R}^2$, the \emph{Frobenius norm approximation}~\cite{SAJJADROBUST,APPROXSYMMPC}, which was originally defined as a pre-conditioner for linear systems~\cite{CIRCAPPROXTOEPLITZ}. The resulting closed-loop properties have not been analysed in detail and the choice of the Frobenius norm has not been justified. The second aim of this paper is to characterise the Frobenius norm approximation.

The final aim of this paper is to propose alternatives to the Frobenius approximation. It will be shown that an approximation based on the Frobenius norm can lead to unstable closed-loop dynamics even when a different structured approximation yields stable dynamics. For this reason, linear matrix inequalities (LMIs) or bilinear matrix inequalities (BMIs) are derived from the stability, performance and robustness properties of the system from Fig.~\ref{fig:fbsystem} for a generic structured approximation $R_\mathcal{S}$. The LMIs or BMIs are then embedded in a semidefinite program (SDP) with the aim of finding a structured approximation that possibly performs better than the Frobenius norm approximation. The SDP can be formulated in the symmetric domain, where the optimisation variable is sparse, which makes this approach suitable for large-scale systems.

We conclude the paper with applying our approach to the electron beam stabilisation problem in a synchrotron light source~\cite{WIEDEMANN,SANDIRAWINDUP,SYNCSYM,SAJJADPADFT,SAJJADROBUST}, which is a particle accelerator that produces high brilliance light by accelerating electrons close to the speed of light. The electron beam dynamics are modelled by~\eqref{eq:CDsystem} with $y(s)$ being the \emph{beam position} and $u(s)$ the setpoints for the \emph{corrector magnets}, whose magnetic fields deflect the electrons. To reduce beam vibrations down to sub-micron level, synchrotrons typically use several hundreds of sensors and actuators that are controlled at a sampling rate $f_s\geq \SI{10}{\kHz}$. At such high sampling rates, the time delay associated with controller computations constitutes up to \SI{20}{\percent} of the total time delay~\cite[Table 2.11.10]{DIAMONDIITDR} and thereby limits the closed-loop bandwidth. Many synchrotrons have various structural symmetries, such as the \emph{block-circulant} and \emph{block-centrosymmetric} symmetry~\cite{SYNCSYM}. Although numerous controllers have been proposed and implemented for the electron beam stabilisation problem~\cite{TAN,APSCOMMISSIONING,FOFBSVD,SANDIRAWINDUP}, only~\cite{SYNCSYM,SAJJADPADFT} have considered (exact) structural symmetry. Exploiting the structural symmetry can significantly reduce the computation time of matrix-vector multiplications by a factor of 12~\cite{SYNCSYM,BCINVADMMCONF}, thereby enabling an increase of the closed-loop bandwidth.

This paper is organised as follows. In Sections~\ref{sec:approx:stability}--\ref{sec:approx:additional}, the stability, performance and robustness properties for the setting from Fig.~\ref{fig:fbsystem} are analysed. In each section, the analysis is followed by deriving LMIs and BMIs that can be embedded in an SDP. In Section~\ref{sec:approx:frobapprox}, the Frobenius norm approximation is analysed, before formulating SDPs for alternative approximations in Section~\ref{sec:approx:otherapprox}. The paper is concluded by applying the results to the electron beam stabilisation problem of a synchrotron, which is an example of a large-scale and high-speed system with cross-directional dynamics.

\emph{Notation and Definitions} Let $\otimes$ denote the Kronecker product and $I_n$ represent the identity matrix in $\R^{n\times n}$. For a scalar, vector or matrix $A$, let $\bar{A}$ denote its complex conjugate, $\herm{A}$ its Hermitian transpose and $\pinv{A}$ its pseudoinverse~\cite[p. 290]{GOLUB4}. Let $\diag(A_1,\dots,A_n)$ denote a diagonal matrix with diagonal elements $A_1,\dots,A_n$. Let $\twonorm{A}$ and $\frobnorm{A}$ denote the spectral and the Frobenius norm, $\onenorm{A}$ and $\infnorm{A}$ the maximum absolute column and row sum, $\kappa(A)\eqdef\twonorm{A}\twonorm{\inv{A}}$ the condition number, and $\rho(A)\eqdef\max_i\abs{\eig_i(A)}$ the spectral radius.
\section{Background: Structural Symmetry\label{sec:background}}
\begin{defn}\label{def:sym}
A set of structurally symmetric matrices is defined as the subspace $\mathcal{S}\eqdef\left\lbrace
\inR{A}{n}{n}
\mid
A\Pi = \Pi A
\right\rbrace$ and associated with a permutation matrix $\inR{\Pi}{n}{n}$, $\Pi_{ij}\in\lbrace 0,1\rbrace$, satisfying $\xTx{\Pi}=I$ and $\Pi^n=I$~\cite[0.9.5]{HORN}. 
\end{defn}
Because $\Pi$ is an orthonormal matrix, there exists $\inC{V}{n}{n}$ with $\herm{V}V=I$ that diagonalises $\Pi$~\cite[Ch.\ 2.5]{GOLUB4}. It holds that $A\in\mathcal{S}$ iff $\herm{V}AV$ is diagonal~\cite[Thm.\ 1.3.12]{HORN}. The matrices in $\mathcal{S}$ form a commutative algebra, i.e.\ $A+B\in\mathcal{S}$, $AB\in\mathcal{S}$ and $AB=BA$ iff $A,B\in\mathcal{S}$. In addition, $\inv{A}\in\mathcal{S}$ if $A\in\mathcal{S}$ is invertible.
\begin{defn}\label{def:compl}
The orthogonal complement $\mathcal{S}^\perp$ of $\mathcal{S}$ is $\mathcal{S}^\perp\eqdef\left\lbrace
\inR{B}{n}{n}\mid\trace (\trans{B}A) = 0 \quad\forall A\in\mathcal{S}\right\rbrace.$
\end{defn}
From $\trace (\trans{B}A)=\trace ((\herm{V}\trans{B}V)(\herm{V}AV))$, it follows that $B\in\mathcal{S}^\perp$ iff $\herm{V}BV$ is \emph{hollow}, i.e.\ a matrix with zero diagonal elements. Since $\herm{V}BV$ and $B\in\mathcal{S}^\perp$ are similar, it holds that $\trace(B) = 0$.
\begin{lem}
A matrix $\inR{C}{n}{n}$ can be uniquely represented as $C=A+B$ with $A\in\mathcal{S}$ and $B\in\mathcal{S}^\perp$.
\end{lem}
\begin{pf}
This follows from computing $\herm{V}C V\!=\!\herm{V}\!AV\!+\!\herm{V}\!BV$ and noting that $B\!\in\!\mathcal{S}^\perp$ iff $\herm{V}BV$ is hollow.\qed
\end{pf}
The concept of structural symmetry can be straightforwardly extended to \emph{block-structural symmetry} by considering matrices $\inR{A}{n_y}{n_u}$ that satisfy $A(\Pi\otimes I_{b_u})=(\Pi\otimes I_{b_y})A$. The matrix $(\herm{V}\otimes I_{b_y})A(V\otimes I_{b_u})$ and $(\herm{V}\otimes I_{b_y})B(V\otimes I_{b_u})$ are then block-diagonal and block-hollow, respectively. In the following, a ``scalar'' symmetry from Def.~\ref{def:sym} will not be distinguished from a block symmetry, e.g.\ writing $R\in\mathcal{S}$ for $\inR{R}{n_y}{n_u}$ may imply that $\exists\, n, b_y, b_u\in\mathcal{Z}_{++}$ such that $n=n_y/b_y=n_u/b_u$ and $R(\Pi\otimes I_{b_u})=(\Pi\otimes I_{b_y})R$ with $\inR{\Pi}{n}{n}$ being associated with $\mathcal{S}$. The following Lemma~\ref{thm:pinv} will be needed for rectangular matrices.
\begin{lem}\label{thm:pinv}
For $\inR{A}{n_y}{n_u}$, $\pinv{A}\in\mathcal{S}$ if $A\in\mathcal{S}$.
\end{lem}
\begin{pf}
Suppose that $A\in\mathcal{S}$, so that $\invbr{\xTx{A}+\delta I}\allowbreak\in\allowbreak\mathcal{S}$ $\forall\delta\in\R$. Using the limit definition of $\pinv{A}$~\cite[P5.5.2]{GOLUB4}: 
\begin{align*}
\pinv{A}\Pi&=\lim_{\delta\rightarrow 0}\invbr{\xTx{A}+\delta I}\trans{A}\Pi\,\overset{\text{Def.~\ref{def:sym}}}{=}\,\Pi\lim_{\delta\rightarrow 0}\invbr{\xTx{A}+\delta I}\trans{A}=\Pi\pinv{A}.
\end{align*}\qed
\end{pf}
\section{Nominal Stability\label{sec:approx:stability}}
To analyse the nominal stability, the approximation $R_\mathcal{S}$ is substituted for $R$ in the model path of Fig.~\ref{fig:fbsystem} and $Q(s)$ is formed using~\eqref{eq:Q}, while assuming that $R$ and $g(s)$ accurately model the CD process. 

Using Fig.~\ref{fig:fbsystem} and defining $P_\mathcal{S}(s)\eqdef R_\mathcal{S} g(s)$, the transfer function from $d(s)$ to $u(s)$ is derived as
\begin{align}
u(s)&=-\invbr{I+Q(s)\br{P\br{s}-P_\mathcal{S}(s)}}Q(s)d(s),\nonumber\\
&=-Q(s)(I+\underbrace{\br{P(s)-P_\mathcal{S}(s)}Q(s)}_{=\Delta_\mathcal{S}R_\mathcal{S}^\dagger T(s)})^{\sm 1}d(s),\nonumber\\
&=-Q(s)\invbr{I+\Phi_\mathcal{S}T(s)}d(s),\label{eq:uCL}
\end{align}
where the push-through rule was used~\cite[Ch.\ 3.2]{SKOGESTADMULTI} and the \emph{error matrix} $\inR{\Phi_\mathcal{S}}{n_y}{n_y}$ defined as
\begin{align}\label{eq:Phi}
\Phi_\mathcal{S}\eqdef\Delta_\mathcal{S} R_\mathcal{S}^\dagger.
\end{align}
After substituting~\eqref{eq:uCL} in the CD system~\eqref{eq:CDsystem}, the closed-loop transfer function from $d(s)$ to $y(s)$ is obtained as
\begin{align}
y(s) &= \left(I-P(s)Q(s)\invbr{I+\Phi_\mathcal{S}T(s)}\right)d(s),\nonumber\\
&= \left(I-T(s) (R_\mathcal{S}+\Delta_\mathcal{S})R_\mathcal{S}^\dagger \invbr{I+\Phi_\mathcal{S}T(s)}\right)d(s),\nonumber\\
&= \underbrace{\left(I-T(s)(I+\Phi_\mathcal{S})\invbr{I+\Phi_\mathcal{S}T(s)}\right)}_{\reqdef S(s)}d(s),\label{eq:closedloop}
\end{align}
where $R_\mathcal{S}R_\mathcal{S}^\dagger=I$ because $n_y\leq n_u$ and $S(s)$ is the \emph{output sensitivity}. If $Q(s)$ is formed using~\eqref{eq:Q} with $R_\mathcal{S}=R$ ($\Phi_\mathcal{S}=0$), then the standard IMC closed-loop~\cite[Ch.\ 4.2]{morari1989robust} is recovered as 
\begin{align}\label{eq:closedloopstandard}
y(s)=(1-T(s))d(s),
\end{align}
which is stable if $T(s)$ is stable. It holds that $y(0)=0$ in both~\eqref{eq:closedloop} and~\eqref{eq:closedloopstandard} if $T(0)=1$, from which it follows that the standard feedback equivalent of Fig.~\ref{fig:fbsystem} implements $n_y$ integrators for any $R_\mathcal{S}$. However, substituting $R_\mathcal{S}$ for the original $R$ introduces an approximation error $\Delta_\mathcal{S}$, which in turn introduces the term $\invbr{I+\Phi_\mathcal{S}T(s)}$ in~\eqref{eq:closedloop} that can be a source of instability. This is investigated in Thm.~\ref{thm:stabdet}.
\begin{thm}\label{thm:stabdet}
The system from Fig.~\ref{fig:fbsystem} with $Q(s)$ defined as in~\eqref{eq:Q} is (internally) stable iff the Nyquist plot of $\det(I+\Phi_\mathcal{S}T(s))=\prod_i(1+\phi_iT(s))$, where $\phi_i$ are the eigenvalues of $\Phi_\mathcal{S}$, does not encircle the origin.
\end{thm}
\begin{pf}
No pole-zero cancellations with $\rre(s)>0$ occur when forming the closed-loop transfer functions~\eqref{eq:uCL} and ~\eqref{eq:closedloop}, which are products of stable transfer functions with $\invbr{I+\Phi_\mathcal{S}T(s)}$. According to the Nyquist stability criterion~\cite[Thm.\ 4.9]{SKOGESTADMULTI}, $\invbr{I+\Phi_\mathcal{S}T(s)}$ is stable iff the Nyquist plot of $\det(I+\Phi_\mathcal{S}T(s))$ does not encircle the origin.\qed
\end{pf}
\noindent Thm.~\ref{thm:stabdet} allows the stability of the system from Fig.~\ref{fig:fbsystem} to be linked to the eigenvalues of the error matrix $\Phi_\mathcal{S}$. In Cor.~\ref{thm:stabdiag}, Thm.~\ref{thm:stabdet} is further simplified.
\begin{cor}\label{thm:stabdiag}
Suppose that $\inv{W}\Phi_\mathcal{S}W=\diag(\phi_1,\dots,\allowbreak\phi_{n_y})$, $\phi_i\in\C$ and $\inC{W}{n_y}{n_y}$. Then the system from Fig.~\ref{fig:fbsystem} is stable iff for each $i=1,\dots,n_y$, none of the Nyquist plots of $1+\phi_iT(s)$ encircles the origin.
\end{cor}
\begin{pf}
The claim follows from diagonalising $\invbr{I+\Phi_\mathcal{S}T(s)}$ and applying the Nyquist stability criterion to the decoupled system. \qed
\end{pf}
\noindent If all eigenvalues of $\Phi_\mathcal{S}$ were real, then, according to Cor.~\ref{thm:stabdiag}, the range of $\phi_i$ that yields a stable system could be computed from the gain margin of $T(s)$. However, since $\Phi_\mathcal{S}\neq\Phi_\mathcal{S}^*$ in general, it must be assumed that some $\phi_i$ are complex-valued or that $\Phi_\mathcal{S}$ is not diagonalisable. A more tractable but conservative condition than Cor.~\ref{thm:stabdiag} is given in Cor.~\ref{thm:stabrho}~\cite[Thm.\ 4.11]{SKOGESTADMULTI}.
\begin{cor}\label{thm:stabrho}
The system from Fig.~\ref{fig:fbsystem} is stable if the spectral radius $\rho(\Phi_\mathcal{S})\eqdef \max_i\abs{\phi_i}$ satisfies $\rho(\Phi_\mathcal{S})<1$, where $\phi_i$ are the eigenvalues of $\Phi_\mathcal{S}$.
\end{cor}
\noindent Cor.~\ref{thm:stabrho} could also be obtained from applying standard techniques from robust control~\cite[Ch.\ 8]{SKOGESTADMULTI}. Note that Thm.~\ref{thm:stabdet} and Corollaries~\ref{thm:stabdiag} and~\ref{thm:stabrho} can also be formulated in the symmetric domain, i.e.\ by substituting $\hat{\Phi}_\mathcal{S}$ for $\Phi_\mathcal{S}$, where
\begin{align}\label{eq:phihat}
\hat{\Phi}_\mathcal{S} \eqdef (V^*\otimes I_{b_y})\Phi_\mathcal{S}(V\otimes I_{b_y}) = \hat{\Delta}_\mathcal{S}\pinv{\hat{R}}_\mathcal{S},
\end{align}
and $\hat{\Delta}_\mathcal{S}\eqdef(\herm{V}\otimes I_{b_y})\Delta_\mathcal{S}(V\otimes I_{b_u})$.
\subsection{Stability Conditions\label{sec:stabconds}}
With the controller $Q(s)$ being fixed as in~\eqref{eq:Q}, the nominal stability conditions depend on $T(s)$ and the choice of $R_\mathcal{S}$. For a given $R_\mathcal{S}$, if the system is unstable, one possibility would be to substitute $\alpha T(s)$, $0< \alpha < 1$, for $T(s)$, i.e.\ reducing the steady-state gain of $Q(s)$. However, according to~\eqref{eq:closedloop} and~\eqref{eq:closedloopstandard}, this would result in $y(0)\neq 0$ and therefore introduce an undesirable steady-state error. 

Alternatively, the spectral radius of $\Phi_\mathcal{S}$ can be upper-bounded using matrix inequalities, which can subsequently be used to choose an approximation $R_\mathcal{S}$ that gives a favourable spectral radius of $\Phi_\mathcal{S}$. For that purpose, $\Phi_\mathcal{S}$ can be expanded as
\begin{align}\label{eq:Phisimple}
\Phi_\mathcal{S}
=\Delta_\mathcal{S}R_\mathcal{S}^\dagger
\overset{\eqref{eq:Rsdef}}{=}(R-R_\mathcal{S})R_\mathcal{S}^\dagger=RR_\mathcal{S}^\dagger-I,
\end{align}
where it is assumed that $R_\mathcal{S}R_\mathcal{S}^\dagger=I$ because $n_y\leq n_u$. Using~\eqref{eq:Phisimple}, any upper bound on $\Phi_\mathcal{S}$ can be formulated in terms of $R_\mathcal{S}^\dagger$ or, after mapping~\eqref{eq:Phisimple} to symmetric domain, in terms of $\hat{R}_\mathcal{S}^\dagger$. 
\subsubsection{Upper Bound via 2-norm}
The spectral radius $\rho(\Phi_\mathcal{S})$ can be upper-bounded by ~\cite[Thm.\ 5.6.14]{HORN}
\begin{align}\label{eq:otherbound}
\rho(\Phi_\mathcal{S})\leq \twonorm{\Phi_\mathcal{S}^k}^{1/k},\qquad k\in\mathbb{Z}_{++},
\end{align}
with $\lim_{k\rightarrow\infty}\twonorm{\Phi_\mathcal{S}^k}^{1/k}=\rho(\Phi_\mathcal{S})$. Although~\eqref{eq:otherbound} uses the 2-norm, any other sub-multiplicative norm, such as the Frobenius norm, could also be used. By choosing $k=1$ in~\eqref{eq:otherbound} and substituting the right-hand side of~\eqref{eq:Phisimple}, a sufficient condition for nominal stability is $\twonorm{RR_\mathcal{S}^\dagger-I}<1$, which, using the Schur complement~\cite[Ch.\ 2]{BOYDLMI}, can be reformulated as the following linear matrix inequality (LMI):
\begin{align}\label{eq:NS1}\tag{NS1}
\begin{bmatrix}
I & RX-I \\ (RX-I)^* & I
\end{bmatrix}\succ 0,
\end{align}
where $X\eqdef R_\mathcal{S}^\dagger\in\mathcal{S}$ (cf. Lemma~\ref{thm:pinv}). 

Because $\twonorm{A^2}^{1/2}\leq\twonorm{A}$, a possibly tighter bound can be obtained by choosing $k=2$ in~\eqref{eq:otherbound}, which yields the sufficient stability condition $\twonorm{(RX-I)^2}<1$. Using the Schur complement, this can be reformulated as
\begin{align}\label{eq:NS2}\tag{NS2}
\begin{bmatrix}
I & RXRX-2 RX +I\\
\hermbr{\dots} & I
\end{bmatrix}
\succ 0.
\end{align}
Constraint~\eqref{eq:NS2} is a \emph{bilinear matrix inequality} (BMI) in $X$ that can be solved using different convexifying techniques. One possibility is to use the approach presented in~\cite{CONVEXBMI} (Section~\ref{sec:approx:BMI}), which finds a solution to an optimisation problem with BMI constraints by solving a sequence of SDPs. Note that~\eqref{eq:NS2} is never more conservative than~\eqref{eq:NS1}.
\subsubsection{Lyapunov Certificate}
The problem of finding $X$ such that $\rho(RX-I)<1$ can be recast using a discrete-time Lyapunov function approach~\cite[Ch.\ 1.4.4]{ADVLMI}. It holds that $\rho(RX-I)<1$ iff there exists $P\in\mathbb{S}_{++}$, where $\mathbb{S}_{++}$ is the set of real symmetric positive definite matrices, such that $P-\herm{(RX-I)} P (RX-I)\succ 0$~\cite[Ch.\ 1.4.4]{ADVLMI}. Applying the Schur complement to this matrix inequality leads to the following constraint:
\begin{equation}\label{eq:NS3}\tag{NS3}
\begin{bmatrix}\inv{P} & RX-I\\\hermbr{RX-I} & P\end{bmatrix}\succ 0,
\end{equation}
which, after pre- and post-multiplication with $\diag(I,\allowbreak\inv{P})$, can be interpreted as a BMI in $X$ and $\inv{P}$.

In contrast to the constraints~\eqref{eq:NS1} and~\eqref{eq:NS2}, constraint~\eqref{eq:NS3} introduces a dense matrix variable $P$ and eventually becomes difficult to solve for large-scale matrices. Alternatively, one can fix $P$ to have the same structural symmetry as $X$ and reformulate~\eqref{eq:NS3} as
\begin{equation}\label{eq:NS4}\tag{NS4}
\begin{bmatrix}
\inv{P_\mathcal{S}} & RZ_\mathcal{S}-\inv{P_\mathcal{S}}\\
\hermbr{RZ_\mathcal{S}-\inv{P_\mathcal{S}}} & \inv{P_\mathcal{S}}
\end{bmatrix}\succ 0,
\end{equation}
where $P_\mathcal{S}\in\mathcal{S}$ and $Z_\mathcal{S}\eqdef X \inv{P_\mathcal{S}}\in\mathcal{S}$. Constraint~\eqref{eq:NS4} is an LMI in $Z_\mathcal{S}$ and $\inv{P_\mathcal{S}}$. Note that if an $X$ is found that satisfies~\eqref{eq:NS1}, then the same $X$ satisfies~\eqref{eq:NS3} or~\eqref{eq:NS4} with $P=P_\mathcal{S}=I$, i.e.~\eqref{eq:NS3} and~\eqref{eq:NS4} are never more conservative than~\eqref{eq:NS1}.
\section{Nominal Performance\label{sec:approx:performance}}
In order to measure the impact of an approximation $R_\mathcal{S}$ on the performance, the output of the system that uses $R$, $y(s)$, can be compared with $y_\mathcal{S}(s)$, the output of the one that uses $R_\mathcal{S}$. Subtracting~\eqref{eq:closedloop} from~\eqref{eq:closedloopstandard}, the error $e(s)\eqdef y(s)-y_\mathcal{S}(s)$ is $e(s)=E(s)d(s)$ with
\begin{align}\label{eq:error}
E(s)&\eqdef T(s)\left((I+{\Phi}_\mathcal{S})\invbr{I+T(s){\Phi}_\mathcal{S}} -I\right)
= T(s)\left(1-T(s)\right)\Phi_\mathcal{S}\invbr{I+T(s){\Phi}_\mathcal{S}},
\end{align}
where $E(0)=\lim_{\omega\rightarrow\infty}E(\mathrm{j}\omega)=0$. An approximation $R_\mathcal{S}$ that minimises $\twonorm{E(s)}$ therefore yields a similar closed-loop response to a system that uses $R$.
\subsection{Performance Bounds\label{sec:constraint}}
To obtain a more tractable form than~\eqref{eq:error}, the term ${\Phi}_\mathcal{S}\invbr{I+T(s){\Phi}_\mathcal{S}}$ is expanded using the Neumann series~\cite[Ch.\ 5.6, P26]{HORN} as
\begin{align}
{\Phi}_\mathcal{S}\invbr{I+T(s){\Phi}_\mathcal{S}} &={\Phi}_\mathcal{S}\sum_{k=0}^\infty(-T(s){\Phi}_\mathcal{S})^k
={\Phi}_\mathcal{S}-T(s){\Phi}_\mathcal{S}^2+\mathcal{O}({\Phi}_\mathcal{S}^3),\label{eq:neumann}
\end{align}
where it is assumed that $\rho(T(\mathrm{j}\omega)\Phi_\mathcal{S})<1$. Combining~\eqref{eq:error} and~\eqref{eq:neumann}, the magnitude of $E(s)$ can be upper-bounded by
\begin{equation}\label{eq:npbound}
\begin{aligned}
\twonorm{E(s)} \leq \abs{ T(s)(1-T(s))}&\twonorm{{\Phi}_\mathcal{S}-T(s){\Phi}_\mathcal{S}^2}+\twonorm{\mathcal{O}({\Phi}_\mathcal{S}^3)}.
\end{aligned}
\end{equation}
Ignoring higher-order terms in~\eqref{eq:npbound}, $R_\mathcal{S}^\dagger$ can be chosen to minimise an upper bound $\sqrt{\alpha_\omega}\in\R_{++}$ on $\twonorm{{\Phi}_\mathcal{S}-T(\jw){\Phi}_\mathcal{S}^2} = \twonorm{RR_\mathcal{S}^\dagger-I-T(\jw)(RR_\mathcal{S}^\dagger-I)^2}$ at a particular frequency $\omega$, which can be formulated using the Schur complement as
\begin{align}\label{eq:NP1}\tag{NP1}
\begin{bmatrix}I & RX - I-T(\mathrm{j}\omega)(RX - I)^2\\ \hermbr{\dots} & \alpha_\omega I\end{bmatrix}\succeq 0,
\end{align}
where $X=R_\mathcal{S}^\dagger$. If ~\eqref{eq:NP1} holds, then $\twonorm{E(\jw)}\leq\sqrt{\alpha_\omega}\abs{ T(\jw)(1-T(\jw))}$. Note that~\eqref{eq:NP1} is a BMI, but in the limit $\omega\rightarrow\infty$ the following LMI is obtained:
\begin{align}\label{eq:NP2}\tag{NP2}
\begin{bmatrix}I & RX - I\\ (RX - I)^* & \alpha_\infty I\end{bmatrix}\succeq 0,
\end{align}
which reduces to the nominal stability condition~\eqref{eq:NS1} for $\alpha_\infty=1$.
\section{Robust Stability\label{sec:approx:additional}}
When the plant $P(s)=Rg(s)$ is approximated using $P_\mathcal{S}(s)=R_\mathcal{S}g(s)$, thereby artificially introducing the approximation error $\Delta_\mathcal{S}$~\eqref{eq:Rsdef}, it is assumed that $P(s)$ is known. In this section, it is assumed that $P(s)$ has an additional unknown component $\Theta : \C\mapsto \C^{n_y \times n_u}$, i.e.
\begin{align}\label{eq:additional}
P(s)\eqdef (R_\mathcal{S}+\Delta_\mathcal{S})g(s)+\Theta(s),
\end{align}
where $R=R_\mathcal{S}+\Delta_\mathcal{S}$ is known and $R_\mathcal{S}$ is used to obtain the IMC filter $Q(s)$~\eqref{eq:Q}. 

\begin{figure}
\inputtikz{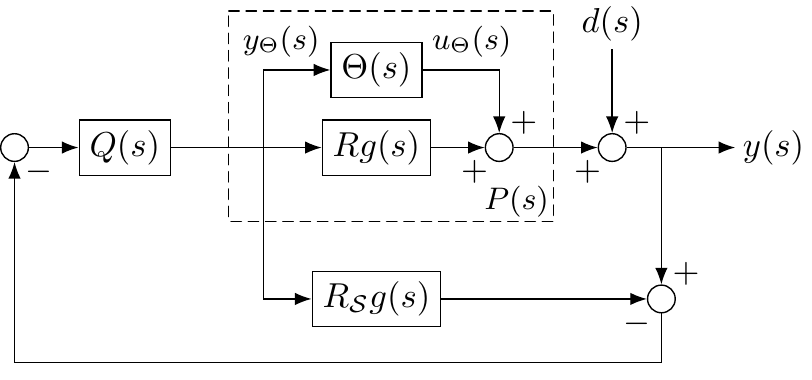}
\caption{IMC structure with unknown uncertainty $\Theta(s)$ and known $R$ and $g(s)$. The model path contains the approximation $R_\mathcal{S}$ that is used to form $Q(s)$.\label{fig:fbsystemrob}}
\end{figure}

It is assumed that a given $R_\mathcal{S}$ yields a stable system for $\Theta(s)=0$. Then, for $\Theta(s)\neq 0$, the system from Fig.~\ref{fig:fbsystemrob} is stable iff~\cite[Thm.\ 8.1]{SKOGESTADMULTI}
\begin{align}\label{eq:detcondrob}
\det (I-M(\jw)\Theta(\jw))\neq 0 \quad\forall\omega,
\end{align}
where $M(s)\eqdef -Q(s)\invbr{I+T(s)\Phi_\mathcal{S}}$ is the transfer function from $u_\Theta(s)$ to $y_\Theta(s)$ that equals the one from $d(s)$ to $u(s)$~\eqref{eq:uCL}. A sufficient condition for~\eqref{eq:detcondrob} is
\begin{align}\label{eq:rhocondrob}
\rho\left(M(\jw)\Theta(\jw)\right)<1\quad\forall\omega,
\end{align}
which, analogous to the \textit{nominal} stability conditions from Section~\eqref{sec:stabconds}, can be upper-bounded using the 2-norm to obtain $\twonorm{\Theta(\jw)}<1/\twonorm{M(\jw)}$, where
\begin{align}\label{eq:normcondrob}
\frac{1}{\twonorm{M(\jw)}}=\frac{\abs{g(\jw)}}{\abs{T(\jw)}\twonorm{R_\mathcal{S}^\dagger\invbr{I+T(\jw)\Phi_\mathcal{S}}}}.
\end{align}
If, for a given uncertainty $\Theta(s)$ condition~\eqref{eq:normcondrob} is satisfied $\forall\omega$, then the system from Fig.~\ref{fig:fbsystemrob} is stable. Moreover, a small $\twonorm{R_\mathcal{S}^\dagger\invbr{I+T(\jw)\Phi_\mathcal{S}}}$ allows for a large $\twonorm{\Theta(\jw)}$.
\subsection{Robustness Bounds\label{sec:robbounds}}
To obtain a robustness condition that can be embedded in an optimisation problem, the right-hand side of~\eqref{eq:normcondrob} is expanded using a Neumann series~\cite[Ch.\ 5.6, P26]{HORN} and~\eqref{eq:Phisimple} as
\begin{align}
R_\mathcal{S}^\dagger\invbr{I+T(s){\Phi}_\mathcal{S}}&=R_\mathcal{S}^\dagger\sum_{k=0}^\infty(-T(s){\Phi}_\mathcal{S})^k=R_\mathcal{S}^\dagger+\mathcal{O}\left(\left(R_\mathcal{S}^\dagger\right)^2\right).\label{eq:neumann2}
\end{align}
An approximation $R_\mathcal{S}$ that yields a robust system therefore tends to make $\twonorm{\pinv{R}_\mathcal{S}}$ small, which is equivalent to decreasing the steady-state gain of the IMC controller. A robust stability condition can be formulated as $\twonorm{R_\mathcal{S}^\dagger}\leq\sqrt{\beta}$ for some $\beta\in\R_{++}$, which can be reformulated using the Schur complement as
\begin{equation}\label{eq:RS}\tag{RS}
\begin{bmatrix}I & X\\\herm{X} & \beta I\end{bmatrix}\succeq 0.
\end{equation}
\section{Approximations with Structural Symmetries\label{sec:approx:approx}}
\subsection{Frobenius Norm Approximation\label{sec:approx:frobapprox}}
Approximations of the form 
\begin{align}\label{eq:genapprox}
R_\mathcal{S}^{(\cdot)} = \argmin_{X\in\,\mathcal{S}} \norm{X- R}^2_{(\cdot)},
\end{align}
where $(\cdot)=\lbrace\text{F},1,\infty\rbrace$, have been proposed in several applications~\cite{CIRCAPPROX,SAJJADPADFT,SAJJADROBUST,PHDDANIELSON,APPROXSYMMPC}. In~\cite{SAJJADPADFT,SAJJADROBUST}, the Frobenius norm is used and applied to a  synchrotron orbit feedback control problem, but the closed-loop properties are not related to the choice of the approximation nor is the choice of the Frobenius norm justified. In~\cite{APPROXSYMMPC}, the 1-norm and the Frobenius norm are applied to obtain approximations used in a robust model predictive control problem. However, in none of the applications has it been noted that considering
\begin{align}\label{eq:anynorm}
\anynorm{\Phi_\mathcal{S}}\leq\anynorm{\Delta_\mathcal{S}}\anynorm{R_\mathcal{S}^\dagger}=\anynorm{R-R_\mathcal{S}}\anynorm{R_\mathcal{S}^\dagger},
\end{align}
where $\anynorm{\cdot}$ is an arbitrary sub-multiplicative norm, it becomes clear that an approximation of the form~\eqref{eq:genapprox} minimises the upper bound $\anynorm{R_\mathcal{S}-R}$ on the error matrix $\norm{\Phi_\mathcal{S}}$. The term $\anynorm{R_\mathcal{S}-R}$ can be interpreted as a first-order approximation of the nominal stability condition~\eqref{eq:otherbound} and performance bounds~\eqref{eq:neumann}, and the term $\anynorm{R_\mathcal{S}^\dagger}$ as a first-order approximation of the robust stability condition~\eqref{eq:normcondrob}.

Even though the matrix norms $(\cdot)=\lbrace\text{F},1,2,\infty\rbrace$ are \textit{equivalent}~\cite[Ch.\ 2.3.2]{GOLUB4}, it is unclear which choice of norm in~\eqref{eq:genapprox} yields the best results. However, when the Frobenius norm is used the approximation error $\Delta_\mathcal{S}$ inherits a special structure that is characterised in Lemma~\ref{thm:delta}:
\begin{lem}\label{thm:delta}
If $R_\mathcal{S}$ is obtained from~\eqref{eq:genapprox} with $(\cdot)=\text{F}$, then $\Delta_\mathcal{S}^\text{F}\eqdef R-R_\mathcal{S}^\text{F}\in\mathcal{S}^\perp$ and $\hat{\Delta}_\mathcal{S}^\text{F}\eqdef (V^*\otimes I_{b_y})\Delta_\mathcal{S}^\text{F}(V\otimes I_{b_u})$ is (block-)hollow.
\end{lem}
\begin{pf}
Because the Frobenius norm is invariant to pre- and post-multiplication with orthogonal matrices~\cite[Ch.\ 2.3.5]{GOLUB4}, problem~\eqref{eq:genapprox} can be reformulated for $(\cdot)=\text{F}$ as $\hat{R}_\mathcal{S}^\text{F} = \argmin_{\hat{X}\in \herm{V} \mathcal{S} V}\frobnorm{\herm{V}RV-\hat{X}}$, where $\hat{X}$ is diagonal. The minimum is attained when $\hat{X}$ equals the diagonal part of $\herm{V}RV$ and according to Def.~\ref{def:compl}, $\hat{\Delta}_\mathcal{S}\in\herm{V} \mathcal{S}^\perp V$ is hollow. The extension to block-structural symmetries is analogous.\qed
\end{pf}
\noindent As a consequence of Lemma~\ref{thm:delta} and the block-diagonal property of $\hat{R}_\mathcal{S}$, it follows that $\hat{\Phi}_\mathcal{S}=\hat{\Delta}_\mathcal{S}\hat{R}_\mathcal{S}^\dagger$ is block-hollow too. Suppose that the \emph{original} matrix $R$ is mapped to the symmetric domain, giving $\hat{R}=(\herm{V}\otimes I_{b_y})R(V\otimes I_{b_u})$, and then partitioned as
\begin{align}\label{eq:structure}
\hat{R} = \hat{R}_\mathcal{S}^\text{F} + \hat{\Delta}_\mathcal{S}^\text{F}=&
\begin{bmatrix}
\hat{r}_{\mathcal{S},1} \\ & \ddots\\ & & \hat{r}_{\mathcal{S},n}
\end{bmatrix}
+\begin{bmatrix}
0 & \hat{\delta}_{12} & \dots & \hat{\delta}_{1n}\\[-0.5em]
\hat{\delta}_{21} & 0 &  & \vdots\\[-0.5em]
\vdots            &  &  \ddots & \hat{\delta}_{(n\sm 1)n}\\[-0.5em]
\hat{\delta}_{n1} & \dots  & \hat{\delta}_{n(n\sm 1)} & 0
\end{bmatrix} ,
\end{align}
where $\inC{\hat{r}_{\mathcal{S},i}, \hat{\delta}_{ij}}{b_y}{b_u}$, then the block-hollow property of $\hat{\Phi}_\mathcal{S}$ can be used to apply a Ger\v{s}gorin-circle-type theorem for block-partitioned matrices~\cite{FGOLDVARGA} that relates~\eqref{eq:structure} to the spectral radius $\rho(\hat{\Phi}_\mathcal{S})$, which, considering that~\eqref{eq:phihat} is a similarity transformation, equals $\rho(\Phi_\mathcal{S})$.
\begin{thm}\label{thm:upperbound}
The spectral radius $\rho(\hat{\Phi}_\mathcal{S})$ satisfies $\rho(\hat{\Phi}_\mathcal{S})\leq U$, where
\begin{align*}
U\!\eqdef\min
\Bigg\lbrace  
&\max_{i=1,\dots,n}\,\sum_{\substack{j=1\\ j\neq i}}^n\anynorm{\hat{\delta}_{ij}\hat{r}_{\mathcal{S},i}^\dagger},
\max_{j=1,\dots,n}\,\sum_{\substack{i=1\\ i\neq j}}^n\anynorm{\hat{\delta}_{ij}\hat{r}_{\mathcal{S},i}^\dagger}
 \Bigg\rbrace,
\end{align*}
for block-hollow $\inC{\hat{\Phi}_\mathcal{S}=\hat{\Delta}_\mathcal{S}^\text{F}(\hat{R}_\mathcal{S}^\text{F})^\dagger}{n_y}{n_y}$ partitioned as in~\eqref{eq:structure} and any sub-multiplicative norm $\norm{\cdot}$.
\end{thm}
\begin{pf}
Each eigenvalue $\phi_k$ of $\inC{A}{n_y}{n_y}$ satisfies~\cite[Thm.\ 2]{FGOLDVARGA}
\begin{align*}
\invbr{\anynorm{\invbr{A_{ii}-\phi_k I}}}\leq \sum_{\substack{j=1\\ j\neq i}}^n\anynorm{A_{ij}},
\end{align*}
where $A$ is partitioned into blocks $\inC{A_{ij}}{b_y}{b_y}$. If $A$ is block-hollow, $A_{ii}=0$ and $\invbr{\anynorm{\invbr{A_{ii}-\phi_k I}}}=\abs{\phi_k}$. It remains to substitute $\hat{\Phi}_{\mathcal{S},ij}=\hat{\delta}_{ij}\hat{r}_{\mathcal{S},i}^\dagger$ for $A_{ij}$.\qed
\end{pf}
\noindent Note that the matrices $(\hat{R}_\mathcal{S}^\text{F})^\dagger\hat{\Delta}_\mathcal{S}^\text{F}$ and $\hat{\Delta}_\mathcal{S}^\text{F}(\hat{R}_\mathcal{S}^\text{F})^\dagger$ share the same non-zero eigenvalues~\cite[Ch.\ A.2.1]{SKOGESTADMULTI}, so that Thm.~\ref{thm:upperbound} can also be applied to $(\hat{R}_\mathcal{S}^\text{F})^\dagger\hat{\Delta}_\mathcal{S}^\text{F}$. The following Cor.~\ref{thm:bdom} relates Thm.~\ref{thm:upperbound} to the nominal stability of the closed-loop system through a block-diagonal dominance condition on the partitioning~\eqref{eq:structure}, and is in line with similar results on the decoupling of MIMO systems and decentralised control~\cite[Ch.\ 3.6.2]{SKOGESTADMULTI}; \cite[Ch.\ 4.6]{MACIEJOWSKI}; \cite[Ch.\ 14.4.3]{morari1989robust}.
\begin{cor}\label{thm:bdom}
The system from Fig.~\ref{fig:fbsystem} is nominally stable if $\hat{R}_\mathcal{S}^\text{F}+\hat{\Delta}_\mathcal{S}^\text{F}=\hat{P}(0)$ is strictly column or row {block-diagonally dominant}~\cite[Def. 1]{FGOLDVARGA}, e.g.\ if
\begin{align}\label{eq:bdom}
\invbr{\anynorm{\hat{r}_{\mathcal{S},i}^\dagger}}>\sum_{\substack{j=1\\ j\neq i}}^n\anynorm{\hat{\delta}_{ij}},
\end{align}
for $i=1,\dots,n$ and any sub-multiplicative norm $\norm{\cdot}$.
\end{cor}
\begin{pf} By Cor.~\ref{thm:stabrho}, the system from Fig.~\ref{fig:fbsystem} is nominally stable if $\rho(\Phi_\mathcal{S})=\rho(\hat{\Phi}_\mathcal{S})<1$. From Thm.\ \ref{thm:upperbound}, $\rho(\hat{\Phi}_\mathcal{S})<1$ if $\sum_{j\neq i}\anynorm{\hat{r}_{\mathcal{S},i}^\dagger}\anynorm{\hat{\delta}_{ij}}<1$. Dividing by $\anynorm{\hat{r}_{\mathcal{S},i}^\dagger}$ yields the row-wise block-diagonal dominance condition. The proof is analogous for column block-diagonal dominance.\qed
\end{pf}

The Frobenius norm $\hat{R}_\mathcal{S}^\text{F}$~\eqref{eq:genapprox} yields a block-hollow $\hat{\Delta}_\mathcal{S}^\text{F}$, but it does not necessarily yield the best possible results in terms of stability of the closed-loop system~\eqref{eq:closedloop}. To see this, suppose that the approximation is changed to $\hat{R}_{\mathcal{S}}^\text{F}(1+\alpha)$ with corresponding approximation error $\hat{\Delta}_\mathcal{S}^\text{F}-\alpha\hat{R}_{\mathcal{S}}^\text{F}$ for some scalar $\alpha\in\R_+$. Since $\hat{\Delta}_\mathcal{S}^\text{F}$ is hollow and $\hat{R}_{\mathcal{S}}^\text{F}\perp\hat{\Delta}_\mathcal{S}^\text{F}$, it holds that $\frobnorm{\hat{\Delta}_\mathcal{S}^\text{F}-\alpha\hat{R}_{\mathcal{S}}^\text{F}}\geq\frobnorm{\hat{\Delta}_\mathcal{S}^\text{F}}$, so $\frobnorm{\hat{\Delta}_\mathcal{S}^\text{F}-\alpha\hat{R}_{\mathcal{S}}^\text{F}}$ is not optimal in the sense of~\eqref{eq:genapprox}. The spectral radius condition becomes
\begin{align*}
&\rho\left(\left(\hat{\Delta}_{\mathcal{S}}^\text{F}-\alpha\hat{R}_{\mathcal{S}}^\text{F}\right)\pinvbr{\hat{R}_{\mathcal{S}}^\text{F}(1+\alpha)}\right) =\rho\left(\frac{1}{1+\alpha}\left(
\hat{\Delta}_{\mathcal{S}}^\text{F}\pinv{(\hat{R}_{\mathcal{S}}^\text{F})}-\alpha I
\right)\right)<1.
\end{align*}
If $\phi_{1}\geq\dots\geq\phi_{n_y}$ are the eigenvalues of $\hat{\Delta}_\mathcal{S}^\text{F}\pinv{(\hat{R}_{\mathcal{S}}^\text{F})}$, then $(\phi_i-\alpha)/(1+\alpha)$ are the eigenvalues of $(\hat{\Delta}_{\mathcal{S}}^\text{F}\pinv{(\hat{R}_{\mathcal{S}}^\text{F})}-\alpha I)/(1+\alpha)$. The spectral radius induced by the Frobenius norm approximation can therefore be reduced by choosing a sufficiently small $\alpha$ satisfying $\abs{\phi_1-\alpha}>\abs{\phi_{n_y}-\alpha}$. Such an $\alpha$ always exists if $\abs{\phi_1}\neq\abs{\phi_{n_y}}$.
\begin{rem}
As a consequence of Lemma~\eqref{thm:delta}, the Frobenius approximation can be obtained from the diagonal blocks of $\hat{R}$ instead of solving the SDP~\eqref{eq:genapprox}.
\end{rem}
\subsection{Approximations from SDPs\label{sec:approx:otherapprox}}
In Section~\ref{sec:approx:frobapprox}, it has been shown that the Frobenius norm approximation is possibly sub-optimal with respect to the spectral radius condition of Cor.~\ref{thm:stabrho}. In fact, the Frobenius norm minimises one part of the upper bound~\eqref{eq:anynorm} without considering $\anynorm{\pinv{R}_\mathcal{S}}$. However, for ill-conditioned systems $\anynorm{\pinv{R}_\mathcal{S}}$ might be arbitrarily large and therefore lower the upper bound~\eqref{eq:normcondrob} on the admissible (unknown) uncertainty from the robust stability condition. 
\begin{table}\centering
\caption{Overview of the constraints on nominal stability, nominal performance and robust stability. The second column refers to the equation label.}\label{tab:constraintstable}
\begin{tabular*}{\linewidth}{@{\extracolsep{\fill}} r c c c r @{}}
\toprule
& & Type & \thead{Optimisation\\variables} & \thead{Matrix inequality}\\
\midrule\\[-0.5em]
\multirow{4}{*}[0em]{\begin{sideways}Stability\phantom{asdasssff}\end{sideways}} & \eqref{eq:NS1} & LMI & $X$ & $\begin{bmatrix} I & RX-I \\ \hermbr{\dots} & I \end{bmatrix}\succ 0$\\[1.5em]
& \eqref{eq:NS2} & BMI & $X$ & $\begin{bmatrix} I & (RX-I)^2\\ \hermbr{\dots} & I \end{bmatrix}\succ 0$\\[1.5em]
& \eqref{eq:NS3} & BMI & $X,P$ & $\begin{bmatrix}\inv{P} & RX-I\\\hermbr{\dots} & P\end{bmatrix}\succ 0$\\[1.5em]
& \eqref{eq:NS4} & LMI & $Z_\mathcal{S},\inv{P_\mathcal{S}}$ & $\begin{bmatrix}\inv{P_\mathcal{S}} & RZ_\mathcal{S}-\inv{P_\mathcal{S}}\\\hermbr{\dots} & \inv{P_\mathcal{S}}\end{bmatrix}\succ 0$\\[1.5em]
\midrule\\[-0.5em]
\multirow{2}{*}[0.1em]{\begin{sideways}Performance\end{sideways}} & \eqref{eq:NP1} & BMI & $X,\alpha_\omega$ & $\begin{bmatrix}I & RX - I-T(\mathrm{j}\omega)(RX - I)^2\\ \hermbr{\dots} & \alpha_\omega I\end{bmatrix}\succeq 0$\\[1.5em]
& \eqref{eq:NP2} & LMI & $X,\alpha_\infty$ & $\begin{bmatrix}I & RX - I\\ \hermbr{\dots} & \alpha_\infty I\end{bmatrix}\succeq 0$\\[1.5em]
\midrule\\[-0.5em]
\thead{\begin{sideways}Robustness\end{sideways}} & \eqref{eq:RS} & LMI & $X,\beta$ & $\begin{bmatrix}I & X\\\herm{X} & \beta I\end{bmatrix}\succeq 0$\\[1.5em]
\bottomrule
\end{tabular*}
\end{table}%

The stability, performance and robustness conditions from Sections~\ref{sec:approx:stability}-\ref{sec:approx:additional}, which are summarised in Table~\ref{tab:constraintstable}, can be used to formulate optimisation problems that lead to alternative approximations. For example, consider combining a convex objective function $f:\R_+\times\R_+\mapsto\R_+$ with the constraints~\eqref{eq:NS1}, \eqref{eq:NP2} and~\eqref{eq:RS} into the optimisation problem:
\begin{equation}\label{eq:optimexample}
\begin{array}{ll}
\displaystyle\minimise_{\substack{X\in\mathcal{S},\\ \alpha_\infty,\,\beta\in\,\R_{++}}} & f(\alpha_\infty,\beta)\\[0.25em] \mbox{subject to } & \mbox{\eqref{eq:NS1},\, \eqref{eq:NP2},\, \eqref{eq:RS}},
\end{array}
\end{equation}
which, if a solution exists, returns an approximation $R_\mathcal{S}=\pinv{X}$ that yields a stable closed-loop satisfying performance and robustness bounds~\eqref{eq:NP2} and~\eqref{eq:RS}, respectively.

If the objective function in~\eqref{eq:optimexample} is convex and if the constraints are \textit{linear} matrix inequalities, then~\eqref{eq:optimexample} is an SDP that can be solved using standard scientific software packages. If some of the constraints in~\eqref{eq:optimexample} are bilinear, a sub-optimal solution can be obtained by lower-bounding the BMIs using LMIs~\cite{VANANTWERP2000363}. An approach that has been applied to the BMI from~\eqref{eq:NS3} is given in~\cite{CONVEXBMI} and presented and applied to the remaining BMIs in the following paragraphs.
\subsubsection{Convexifying Algorithm\label{sec:approx:BMI}}
Suppose that the optimisation problem is
\begin{equation}\label{eq:nonconvex}
\begin{array}{ll}
\displaystyle\minimise_{\inR{X}{n_u}{n_y}} & f(X)\\[0.25em] \mbox{subject to } &X\in\Omega_0,
\end{array}
\end{equation}
where $f:\C^{n_u\times n_y}\mapsto \R$ is a convex and first-order differentiable function bounded from below and the constraint set $\Omega_0$ is given by
\begin{equation}
\begin{aligned}
\Omega_0\eqdef\left\lbrace X\in\mathcal{S}\ \left| \ \mathcal{F}(X)\preceq 0,\ F_i(X)\preceq 0,\right.\right.\qquad&\\
\left. i=1,\dots,N\right\rbrace,&
\end{aligned}
\end{equation}
where $\mathcal{F}(X)\preceq 0$ is a BMI and $F_i(X)\preceq 0$ are LMIs. The following Def.~\ref{def:convexmatfunc} introduces the \emph{convexifying potential matrix functional}~\cite{CONVEXBMI} that is used to upper-bound $\mathcal{F}(X)$.
\begin{defn}[{Convexifying potential matrix functional~\cite{CONVEXBMI}}]\label{def:convexmatfunc}
Given a BMI $\mathcal{F}(X)\preceq 0$, the \emph{convexifying potential matrix functional} is a matrix function $G(X,Y)$ that satisfies (i) $G(X,Y)\succeq 0$, (ii) $G(X,X)=0$, and (iii) $\nabla G(X,X)= 0$ $\forall\, X,Y$ and is such that $\mathcal{F}(X)+G(X,Y)\preceq 0$ is an LMI in $X$.
\end{defn}
\noindent For each of the BMI constraints from Table~\ref{tab:constraintstable}, convexifying potential matrix functionals are derived in App.~\ref{app:bmi} and the resulting LMIs, $\mathcal{F}(X)+G(X,Y)\preceq 0$, are listed in Table~\ref{tab:bmitable}. Note that if the LMI $\mathcal{F}(X)+G(X,Y)\preceq 0$ is satisfied for some $X$, then, according to Def.~\ref{def:convexmatfunc}, the BMI $\mathcal{F}(X)\preceq - G(X,Y)\preceq 0$ is satisfied too.
\begin{table}\centering
\caption{Summary of the LMIs, $\mathcal{F}(X)+G(X,X_k)\preceq 0$, that result from upper-bounding the BMI constraints for nominal stability and performance (App.~\ref{app:bmi}). In the last row, $\mathcal{R}$ and $\mathcal{R}_k$ are used as a shorthand for $RX-I$ and $RX_k-I$, respectively.\label{tab:bmitable}}
\begin{tabular*}{\linewidth}{@{\extracolsep{\fill}}  c  r @{}}
\toprule\\[-0.5em]
\eqref{eq:NS2}  & $\begin{bmatrix}
-I & R(2X\!+\!X_kRX_k\!-\!XRX_k\!-\!X_kRX)\!-\!I\\
\hermbr{\dots} & -I
\end{bmatrix}\preceq 0$\\[1.5em]
\eqref{eq:NS3}  & $\begin{bmatrix}\inv{P}_k ( P-2 P_k)\inv{P}_k & -(RX-I)\\ \hermbr{\dots} & -P\end{bmatrix}\preceq 0$\\[1.5em]
\eqref{eq:NP1}  & $\begin{bmatrix}
-I & \mathcal{R}+T(\mathrm{j}\omega)(\mathcal{R}_k^2-\mathcal{R}_k\mathcal{R}-\mathcal{R}\mathcal{R}_k)\\
\hermbr{\dots} & -\alpha_\omega^2 I
\end{bmatrix}\preceq 0$\\[1.5em]
\bottomrule
\end{tabular*}
\end{table}

After convexifying the BMIs, the LMIs from Table~\ref{tab:constraintstable} are embedded in the iterative procedure from Alg.~\ref{alg:convexifying}~\cite[Alg. 1]{CONVEXBMI}. Given a feasible $X_0\in\Omega_0$, Alg.~\ref{alg:convexifying} repeatedly solves an SDP on Line~\ref{alg:convexifying:sdp} to produce iterates $X_{k+1}\in\Omega_k$, where
\begin{equation}\label{eq:Omega_k}
\begin{aligned}
\Omega_k \eqdef \left\lbrace X\in\mathcal{S}\ \left| \ \mathcal{F}(X)+G(X,X_k)\preceq 0,\right.\right.\qquad&\\
\left. F_i(X)\preceq 0,\,\,i=1,\dots,N\right\rbrace&
\end{aligned}
\end{equation}
is updated at every iteration on Line~\ref{alg:convexifying:omega}, and hence guarantees that the BMI is satisfied. The algorithm terminates once $\anynorm{X_{k+1}-X_k}<\varepsilon$, where $\varepsilon>0$ is fixed. If $\mathcal{F}(X)$ is a concave matrix function, Alg.~\ref{alg:convexifying} converges to a local optimum of~\eqref{eq:nonconvex}~\cite[Thm.\ 5]{CONVEXBMI}, which is, as shown in App.~\ref{app:bmi}, only the case for the BMI constraint~\eqref{eq:NS3}.
\begin{algorithm}
\begin{algorithmic}[1]
 \REQUIRE $X_0\in\Omega_0$
 \ENSURE $X^\star\in\mathcal{S}$
 \STATE $k=0$
 \WHILE{$\anynorm{X_{k+1}-X_k}\geq \varepsilon$}
 	\STATE Update $\Omega_k$ using~\eqref{eq:Omega_k}\label{alg:convexifying:omega}
 	\STATE $X_{k+1}=\argmin_{X\in\Omega_k} \,\, f(X)$\label{alg:convexifying:sdp}
 	\STATE $k\mapsfrom k+1$
 \ENDWHILE
\end{algorithmic}
\caption{Convexifying algorithm~\cite{CONVEXBMI} applied to problem~\eqref{eq:nonconvex}.}\label{alg:convexifying}
\end{algorithm}

If the Frobenius norm approximation yields a stable closed-loop, it can be used to initialise Alg.~\ref{alg:convexifying} as $X_0=\pinv{(R_\mathcal{S}^\text{F})}$, but when $R_\mathcal{S}^\text{F}$ yields an unstable closed-loop, an alternative solution is to obtain $X_0$ from the solution to
\begin{equation}\label{eq:optiminit1}
\begin{array}{ll}
\displaystyle\minimise_{\substack{X\in\mathcal{S},\\ P\in\,\mathbb{S}_{++},\\ \sigma\in\,\R_{++}}} & \sigma\\
\mbox{subject to} &
\begin{bmatrix}\sigma\inv{P} & RX-I\\\hermbr{\dots} & P\end{bmatrix}\succ 0,
\end{array}
\end{equation}
which corresponds to the Lyapunov certificate~\eqref{eq:NS3} with an additional variable $\sigma\in\R_{++}$ that is an upper-bound to the spectral radius $\rho(RX-I)$~\cite[Ch.\ 1.4.4]{ADVLMI}. Problem~\eqref{eq:optiminit1} includes a BMI that can be convexified using the procedure from App.~\ref{app:bmi}:
\begin{equation}\label{eq:optiminit2}
\begin{array}{ll}
\displaystyle\minimise_{\substack{X\in\mathcal{S},\\ P\in\,\mathbb{S}_{++},\\ \sigma\in\,\R_{++}}} &\sigma\\
\mbox{subject to} & \begin{bmatrix}\sigma_k(\sigma_k P\!-\!2\sigma P_k) & -P_k(RX\!-\!I)\\ \hermbr{\dots} & -P\end{bmatrix}\!\preceq\! 0.
\end{array}\!
\end{equation}
If Alg.~\ref{alg:convexifying} is applied to~\eqref{eq:optiminit2}, it must be initialised using $X_0\in\mathcal{S}$, $P_0\in\mathbb{S}_{++}$ and $\sigma_0\in\R_{++}$ that satisfy
\begin{align}
\begin{bmatrix}\sigma_0\inv{P_0} & RX_0-I\\\hermbr{\dots} & P_0\end{bmatrix}\succ 0,
\end{align}
which can always be satisfied by choosing $P_0$ and $\sigma_0$ large. If on termination of Alg.~\ref{alg:convexifying} applied to~\eqref{eq:optiminit2} a solution with $\sigma>1$ is obtained, meaning that the approximation yields an unstable closed-loop, the Alg.~\ref{alg:convexifying} either converged to a local optimum or the underlying system does not allow for an approximation that yields a stable closed-loop with the present control approach. In the former case, Alg.~\ref{alg:convexifying} could be repeated using a different initialisation for $P_0$ and $\sigma_0$.
\section{Numerical Examples\label{sec:approx:applications}}
Consider a circulant system of order $n=3$ with $\inR{R}{3}{3}$ given by
\begin{align}\label{eq:example}
R = R_\mathcal{C}^\text{F}+\Delta_\mathcal{C}^\text{F} =
\begin{bmatrix}r_1 & r_2 & r_3\\ r_3 & r_1 & r_2\\ r_2 & r_3 & r_1\end{bmatrix}+
F_3
\begin{bmatrix}0 & \hat{\delta}_1 & \hat{\delta}_1^*\\ \hat{\delta}_2 & 0 & 0\\ \hat{\delta}_2^* & 0 & 0\end{bmatrix}
\herm{F_3},
\end{align}
where $\hat{\delta}_i\in\C$, $\inC{F_3}{3}{3}$ is the discrete Fourier transform matrix~\cite[Ch.\ 1.4.1]{GOLUB4} and $\mathcal{C}$ refers to the circulant symmetry. Right- and left-multiplying~\eqref{eq:example} with $F_3$ and $F_3^*$ yields
\begin{align}\label{eq:split}
\hat{R}=\hat{R}_\mathcal{C}^\text{F}+\hat{\Delta}_\mathcal{C}^\text{F}=
\begin{bmatrix}\hat{r}_1 & 0 & 0\\ 0 & \hat{r}_2 & 0\\ 0 & 0 & \hat{r}_2^*\end{bmatrix}+
\begin{bmatrix}0 & \hat{\delta}_1 & \hat{\delta}_1^*\\ \hat{\delta}_2 & 0 & 0\\ \hat{\delta}_2^* & 0 & 0\end{bmatrix}.
\end{align}
The eigenvalues of $\hat{\Phi}_\mathcal{C}^\text{F}\eqdef\hat{\Delta}_\mathcal{C}^\text{F}(\hat{R}_\mathcal{C}^\text{F})^{-1}$ are
\begin{align}\label{eq:eigval}
&\phi_1^\text{F} = 0, &\phi_{2,3}^\text{F}=\pm\,\sqrt{\frac{\hat{\delta}_1\hat{\delta}_2}{\hat{r}_1\hat{r}_2}+\hermbr{\frac{\hat{\delta}_1\hat{\delta}_2}{\hat{r}_1\hat{r}_2}}}.
\end{align}
For the remainder of Section~\ref{sec:approx:applications}, it is assumed that $T(s)=1/(s+1)$, so that according to Cor.~\ref{thm:stabdiag}, the resulting closed-loop system is stable if $\abs{\phi_i} <1 \,\forall i$, and unstable if $\rre(\phi_i)\leq -1$ for at least one $i$.
\subsection{Unstable Frobenius Norm Approximation\label{sec:unstabfrob}}
Choosing the values of $\hat{R}_\mathcal{C}^\text{F}$ and $\hat{\Delta}_\mathcal{C}^\text{F}$ as $\hat{r}_1=0.1$, $\hat{r}_2=-2+\mathrm{j}$, $\hat{\delta}_1=1+\mathrm{j}0.2$, and $\hat{\delta}_2=-4-\mathrm{j}4$, results in $\phi_3^\text{F}=-2.5<-1$. Note that $\hat{R}$ is \emph{not} diagonally dominant and Cor.~\ref{thm:bdom} is therefore not satisfied. With the aim of obtaining a stable closed-loop, an approximation is obtained from
\begin{equation}\label{eq:optimexampleustab}
\begin{array}{ll}
\displaystyle\minimise_{\substack{X\in\mathcal{C},\\ \alpha_\infty,\,\beta\in\,\R_{++}}} & (\alpha_\infty+\beta)\\[0.25em] \mbox{subject to } & \mbox{\eqref{eq:NS3},\, \eqref{eq:NP2},\, \eqref{eq:RS}},
\end{array}
\end{equation}
where constraint~\eqref{eq:NS3} is a BMI. Problem~\eqref{eq:optimexampleustab} is therefore solved using Alg.~\ref{alg:convexifying} ($\varepsilon=10^{\sm 3}$), which is in turn initialised using~\eqref{eq:optiminit2} that results after 6 iterations in an approximation $\hat{R}_\mathcal{C}=\diag\left(-3.9, -1.1+\mathrm{j}0.5, -1.1-\mathrm{j}0.5\right)$. The spectral radius is $\rho(\hat{\Phi}_\mathcal{C})=0.87$ and the system therefore stable. Using $\hat{R}_\mathcal{C}$, one could proceed with solving~\eqref{eq:optimexample} to improve performance and robustness properties of the approximation.
\subsection{Stable Frobenius Norm Approximation}
\begin{figure}%
\inputtikz{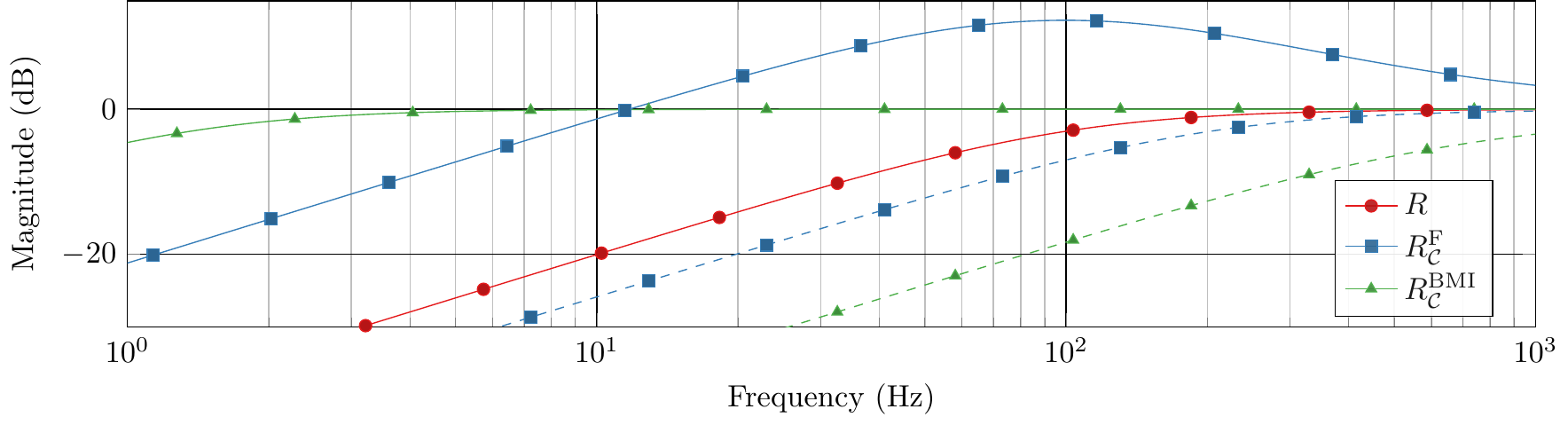}%
\caption{Maximum and minimum (dashed) output sensitivity gains for the stable example from Section~\ref{sec:approx:applications}.\label{fig:stable_frob}}%
\end{figure}

Consider again system~\eqref{eq:split} and the parameters from Section~\ref{sec:unstabfrob}, but divide $\hat{\Delta}_\mathcal{C}^\text{F}$ by $10$, so that $\hat{\delta}_1=0.1+\mathrm{j}0.02$ and $\hat{\delta}_2=-0.4-\mathrm{j}0.4$. Even though $\hat{R}$ is not diagonally dominant, $\rho(\hat{\Phi}_\mathcal{C}^\text{F})=0.25$ and the Frobenius norm approximation is stable. Fig.~\ref{fig:stable_frob} compares the resulting output sensitivity for $Q(s)$~\eqref{eq:Q} formed using $R_\mathcal{C}^\text{F}$ (\addlegendimageintext{mark=square*,draw=legbluedraw,fill=legbluefill}) with the sensitivity for $Q(s)$~\eqref{eq:Q} formed using the original $R$ (\addlegendimageintext{mark=*,draw=legreddraw,fill=legredfill}), where it can be seen that disturbances are amplified by~\SI{12.5}{\dB} at~\SI{100}{\Hz}. To reduce the sensitivity peak, problem~\eqref{eq:optimexample} is initialised using the Frobenius norm approximation and solved using Alg.~\ref{alg:convexifying}. After 15 iterations with constraint~\eqref{eq:NP1} evaluated at $\SI{100}{\Hz}$, Alg.~\ref{alg:convexifying} produces $\hat{R}=\hat{R}_\mathcal{C}^\text{BMI}+\hat{\Delta}_\mathcal{C}^\text{BMI}$ with $\hat{R}_\mathcal{C}^\text{BMI}=\diag(6.8,\,-1+\mathrm{j}0.5,\,-1-\mathrm{j}0.5)$ and
\begin{align*}
\hat{\Delta}_\mathcal{C}^\text{BMI}=
\begin{bmatrix}-6.7 & 0.1+\mathrm{j}0.02 & 0.1-\mathrm{j}0.02\\ -0.4-\mathrm{j}0.4 & -1+\mathrm{j}0.5 & 0\\ -0.4+\mathrm{j}0.4 & 0 & -1-\mathrm{j}0.5\end{bmatrix}.
\end{align*}
The spectral radius is $\rho(\hat{\Phi}_\mathcal{C}^\text{BMI})=0.99$ and the closed-loop system is therefore nominally stable. The resulting output sensitivity is shown in Fig.~\ref{fig:stable_frob} (\addlegendimageintext{mark=triangle*,draw=leggreendraw,fill=leggreenfill}), where it can be seen that the sensitivity peak has been removed. Note that this does not happen at the expense of robustness, because $1/\twonorm{\invbr{\hat{R}_\mathcal{C}^\text{BMI}}}=1.13$ and $1/\twonorm{\inv{(\hat{R}_\mathcal{C}^\text{F})}}=\num{0.1}$.
\section{Case Study: ALBA Synchrotron\label{sec:approx:application}}
The ALBA synchrotron is a synchrotron light source located in Barcelona, Spain, that accelerates electrons to \SI{3}{\GeV}~\cite{ALBACOM}. The electrons circulate around a \SI{270}{\m} circumference \emph{storage ring} that is divided into $n=4$ sections. Each of the four sections has $b_y = 22$ \emph{beam position monitors} (BPMs) and $b_u=22$ corrector magnets, which amounts to a total of $n_y=88$ outputs and $n_u=88$ inputs. The control system, which is typically referred to as the \emph{fast orbit feedback} (FOFB), attenuates vibrations of the electron beam in the horizontal and vertical direction perpendicular to its motion. The horizontal and vertical directions are controlled independently and the following analysis focuses on the vertical direction. At ALBA, the FOFB is sampled at~\SI{10}{\kHz} and designed using the standard feedback structure and a PI controller~\cite{ALBACOMFOFB}. However, the following developments will be based on the (equivalent) IMC structure. The actuator dynamics are modelled as $g(s)=a/(s+a)$ with $a=2\pi\times\SI{700}{\radian\per\second}$ and the complementary closed-loop sensitivity is chosen as $T(s)=\lambda/(s+\lambda)$ with $\lambda=2\pi\times\SI{200}{\radian\per\second}$.

As shown in~\cite{SYNCSYM}, the ALBA \emph{orbit response matrix} $\inR{R}{n_y}{n_u}$ inherits a block-circulant ($\BC$) symmetry and a block-centrosymmetry ($\CS$) from the storage ring structure, but both symmetries are approximate in the sense of~\eqref{eq:approxsymmetry}. These symmetries can be combined into a $\BC\cap\CS$ symmetry, which reduces the computational complexity of the controller further~\cite{SYNCSYM}. Section~\ref{sec:approx:application:frob} presents different approximations of the $\BC\cap\CS$ symmetry and Sections~\ref{sec:application:nomstab}--\ref{sec:application:rob} compare the resulting stability, performance and robustness properties.
\subsection{Approximations\label{sec:approx:application:frob}}
The Frobenius norm approximation is computed using~\eqref{eq:genapprox} for $\mathcal{S}\in\lbrace \BC, \CS, \BC\cap\CS\rbrace$. Table~\ref{tab:specradtableALBA} compares the corresponding spectral radii of $\Phi_\mathcal{S}^\text{F}$ with different metrics of the approximation error $\Delta_\mathcal{S}^\text{F}$. For all symmetries, the approximations satisfy $\rho({\Phi}^\text{F}_\mathcal{S})<1$ and therefore yield stable closed loops.
\begin{table}\centering
\caption{Comparison of the spectral radius $\rho(\Phi_\mathcal{S}^\text{F})$ of the Frobenius norm approximation for the ALBA synchrotron with the approximation error $\Delta_\mathcal{S}^\text{F}$. The metrics in columns 3--5 are computed as $\twonorm{\Delta_\mathcal{S}^\text{F}}/\twonorm{R}$ (2-norm), $\sum_{i,j}\abs{(\Delta_\mathcal{S}^\text{F})_{(i,j)}}/\sum_{i,j}\abs{R_{(i,j)}}$ (mean) and $\max_{i,j}\abs{(\Delta_\mathcal{S}^\text{F})_{(i,j)}}/\max_{i,j}\abs{R_{(i,j)}}$ (max-norm).}%
\label{tab:specradtableALBA}
\begin{tabular*}{\linewidth}{@{}l@{\extracolsep{\fill}} c c c c@{}}
\toprule
& {$\rho(\Phi_\mathcal{S}^\text{F})$} & {\thead{2-norm\\(\%)}} & {\thead{Mean\\(\%)}} & {\thead{Max-norm\\(\%)}} \\
\midrule
$\mathcal{BC}$ & 1.3e-6 & 2.007 & 1.957 & \phantom{0}6.269\\
$\mathcal{CS}$ & 1.5e-1 & 4.234 & 2.849 & 14.998\\
$\mathcal{BC}\cap\mathcal{CS}$ & 1.5e-1 & 4.339 & 3.309 & 15.526\\
\bottomrule
\end{tabular*}%
\end{table}

For the remainder of this section, the analysis is focused onto the combined $\BC\cap\CS$ symmetry, which, according to Table~\ref{tab:specradtableALBA}, results in the largest approximation error and the largest spectral radius. As an alternative to the Frobenius norm approximation, two approximations are derived from the SDPs from Section~\ref{sec:approx:otherapprox}. The ORM of the ALBA synchrotron has $n_y\times n_u=7744$ non-zero elements, which results in large SDPs. In practice, the following problems are therefore formulated in the symmetric domain, where $\hat{R}_{\BC\cap\CS}^{(\cdot)}$ is sparse and has $b_y\times b_u\times (2n-3)/2=1210$ non-zero elements that are purely real or purely imaginary~\cite{SYNCSYM}. The SDPs are solved on a desktop computer (Intel i7-7700 CPU @ \SI{3.1}{\giga\Hz}, \SI{8}{\giga\byte}) using MOSEK~\cite{mosek}.
\subsubsection{Approximation using LMIs}
The LMI constraints~\eqref{eq:NS1},~\eqref{eq:NP2} and~\eqref{eq:RS} from Table~\ref{tab:constraintstable} are combined into the following SDP:
\begin{equation}\label{eq:LMIproblem}
\begin{array}{rlr}\displaystyle
R_{\BC\cap\CS}^\text{LMI}\eqdef	& \displaystyle\argmin_{\substack{X\in\BC\cap\CS,\\ \alpha_\infty,\beta\in\,\R_{++}}}   & \multicolumn{1}{l}{\displaystyle\frac{\alpha_\infty}{\bar{\alpha}_\infty} + \frac{\beta}{\bar{\beta}}} \\
&	\text{subject to} & \quad\mbox{\eqref{eq:NP2},\, \eqref{eq:RS}},\\
&					  & \alpha_\infty < 1,
\end{array}
\end{equation}
where the constraint~\eqref{eq:NS1} is enforced through~\eqref{eq:NP2} with $\alpha_\infty<1$. The objective function, $f(\alpha_\infty,\beta)=\alpha_\infty /\bar{\alpha}_\infty + \beta/\bar{\beta}$, trades off performance versus robustness, and the normalising weights $\bar{\alpha}_\infty$ and $\bar{\beta}$ are chosen as $\bar{\alpha}_\infty\eqdef\twonorm{\Phi_{\BC\cap\CS}^\text{F}}^2$ and $\bar{\beta}\eqdef\twonorm{(R_{\BC\cap\CS}^\text{F})^\dagger}^2$, where $\Phi_{\BC\cap\CS}^\text{F}$ and $R_{\BC\cap\CS}^\text{F}$ stem from the Frobenius norm approximation. When formulated in the symmetric domain, the SDP~\eqref{eq:LMIproblem} is solved within less than a minute. 
\subsubsection{Approximation using BMIs}
With the aim of improving the LMI approximation, problem~\eqref{eq:LMIproblem} is extended with the BMI constraint~\eqref{eq:NP1} to obtain the following non-convex optimisation problem
\begin{equation}\label{eq:BMIproblem}
\begin{array}{rlr}
R_{\BC\cap\CS}^\text{BMI}\eqdef &\displaystyle\argmin_{\substack{X\in\mathcal{S},\\ \alpha_\infty,\alpha_\omega,\beta\in\,\R_{++}}}   & \multicolumn{1}{l}{\displaystyle\frac{\alpha_\infty}{\bar{\alpha}_\infty} + \frac{\beta}{\bar{\beta}} + \frac{\alpha_\omega}{\bar{\alpha}_\omega}} \\
&	\text{subject to} & \mbox{\eqref{eq:NP1},\, \eqref{eq:NP2},\, \eqref{eq:RS}},\\
&					  & \alpha_\infty < 1,
\end{array}
\end{equation}
where $\bar{\alpha}_\omega\eqdef\twonorm{\Phi_\mathcal{S}^\text{F}-T(\mathrm{j}\omega)(\Phi_\mathcal{S}^\text{F})^2}^2$ and constraint~\eqref{eq:NP1} is evaluated at $\omega=2\pi\times \SI{100}{\radian\per\second}$, which will be justified in the following sections. After convexifying the last constraint of~\eqref{eq:BMIproblem} using Table~\ref{tab:bmitable}, problem~\eqref{eq:BMIproblem} can be solved using Alg.~\ref{alg:convexifying}. For the approximation obtained from~\eqref{eq:BMIproblem}, Alg.~\ref{alg:convexifying} was initialised using $R_{\BC\cap\CS}^\text{F}$ and executed 60 iterations before reaching the stopping criteria with $\varepsilon=10^{\sm 3}$, which required \SI{2}{\hour} of computing time. 
\subsection{Nominal Stability\label{sec:application:nomstab}}
Because the optimisation programs~\eqref{eq:LMIproblem} and~\eqref{eq:BMIproblem} enforce closed-loop stability and are solved with no constraint violations, the alternative approximations obtained from~\eqref{eq:LMIproblem} and~\eqref{eq:BMIproblem} both yield stable closed loops. However, in general there is no guarantee that feasible solutions to~\eqref{eq:LMIproblem} and~\eqref{eq:BMIproblem} exist, and because the constraints from Table~\ref{tab:constraintstable} are only sufficient but not necessary, infeasibility of~\eqref{eq:LMIproblem} and~\eqref{eq:BMIproblem} does \emph{not} prove that no stabilising approximation exists.

The spectral radii and additional metrics resulting from~\eqref{eq:LMIproblem} and~\eqref{eq:BMIproblem} are compared with the Frobenius norm approximation in Table~\ref{tab:approxall}, where it can be seen that the spectral radii $\rho(\Phi_{\BC\cap\CS}^{\text{LMI}})$ and $\rho(\Phi_{\BC\cap\CS}^{\text{BMI}})$ are smaller than 1, but over 2 times larger than $\rho(\Phi_{\BC\cap\CS}^\text{F})$. However, the spectral radius is \textit{not} a good measure for robustness, as will be seen in Section~\ref{sec:application:rob}.
\begin{table}\centering
\caption{Comparison of spectral radii $\rho(\Phi_\mathcal{S}^{(\cdot)})$ and 2-norms of $\Phi_\mathcal{S}^{(\cdot)}$, $(R_\mathcal{S}^{(\cdot)})^\dagger$ and $\Delta_\mathcal{S}^{(\cdot)}$ resulting from different approximations for the $\mathcal{S}=\BC\cap\CS$ symmetry at the ALBA synchrotron.}
\label{tab:approxall}
\begin{tabular*}{\linewidth}{@{}l@{\extracolsep{\fill}} c c c c @{}}
\toprule
 & $\rho(\Phi_\mathcal{S}^{(\cdot)})$ & $\twonorm{\Delta_\mathcal{S}^{(\cdot)}}$ & $\twonorm{\Phi_\mathcal{S}^{(\cdot)}}$ & $\twonorm{(R_\mathcal{S}^{(\cdot)})^\dagger}$\\
\midrule
$R_{\BC\cap\CS}^\text{F}$ & 0.1503 & 1.1414 & 0.4736 & 26.8679\\[.25em]
$R_{\BC\cap\CS}^\text{LMI}$ & 0.5259 & 9.4590 & 0.5838 & 18.9985\\[.25em]
$R_{\BC\cap\CS}^\text{BMI}$ & 0.4156 & 1.4024 & 0.4315 & 17.8677\\
\bottomrule
\end{tabular*}
\end{table}
%
\subsection{Nominal Performance\label{sec:application:nomperf}}
\begin{figure}
\inputtikz{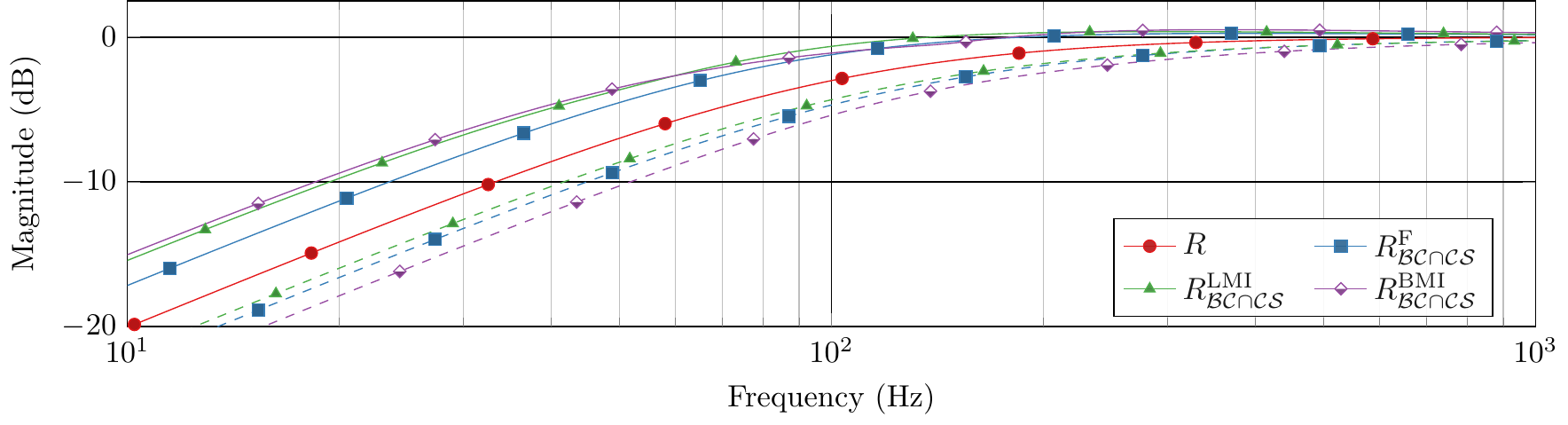}%
\caption{Maximum and minimum (dashed) sensitivity gains for the original system and the $\BC\cap\CS$ approximations of the ALBA synchrotron.}\label{fig:sensitivity}
\end{figure}

The maximum and minimum (dashed) output sensitivity gains~\eqref{eq:closedloop} of the system from Fig.~\ref{fig:fbsystem} are shown in Fig.~\ref{fig:sensitivity} for $Q(s)$ formed using the original $R$ (\addlegendimageintext{mark=*,draw=legreddraw,fill=legredfill}) and the approximations from Section~\ref{sec:approx:application:frob}. For the original $R$, the minimum and the maximum sensitivity gains coincide, which is a consequence of~\eqref{eq:closedloopstandard}. As is typical for synchrotrons~\cite{SANDIRAWINDUP}, the control problem is aggravated by the large condition number $\kappa(R)=856$. In practice, this is considered by introducing a \emph{regularisation matrix} in the feedback path of Fig.~\ref{fig:fbsystem}, causing a difference in bandwidth between the minimum and the maximum sensitivity gains.

Compared to the original system, the closed-loop bandwidth of the systems that are controlled using approximations is lowered by \SI{100}{\Hz}. Measured by the maximum sensitivity gain, the Frobenius norm approximation $R_{\BC\cap\CS}^\text{F}$ (\addlegendimageintext{mark=square*,draw=legbluedraw,fill=legbluefill}) performs best and produces a low-frequency attenuation that is roughly \SI{3}{\dB} higher than the low-frequency attenuation of the original system. The approximation $R_{\BC\cap\CS}^\text{LMI}$ (\addlegendimageintext{mark=triangle*,draw=leggreendraw,fill=leggreenfill}) performs worse than $R_{\BC\cap\CS}^\text{F}$ and produces a worst-case low-frequency disturbance attenuation that is roughly \SI{5}{\dB} higher than the low-frequency attenuation of the original system.

With the aim of reducing the performance difference produced by $R_{\BC\cap\CS}^\text{LMI}$, the SDP problem~\eqref{eq:LMIproblem} was extended with the BMI constraint~\eqref{eq:NP1} evaluated at $\SI{100}{\Hz}$ to obtain $R_{\BC\cap\CS}^\text{BMI}$ (\addlegendimageintext{mark=halfdiamond*, draw=legvioletdraw,fill=legvioletfill}). In Fig.~\ref{fig:fbsystem}, it can be seen that the addition of the BMI constraint lowers the maximum sensitivity gain by roughly \SI{1}{\dB} at $\SI{100}{\Hz}$. Additional BMI constraints at different frequencies could be integrated in~\eqref{eq:BMIproblem} to further reduce the performance difference.

In all cases, the approximations also affect the minimum output sensitivity gain, and according to Fig.~\ref{fig:sensitivity}, perform better for certain disturbance directions. However, the strong directionality of the system also affects the disturbance spectrum, which is more pronounced for directions associated with large singular values of $R$, and a detailed performance analysis therefore requires to consider the disturbance spectrum at the ALBA synchrotron.
%
\subsection{Robust Stability\label{sec:application:rob}}
\begin{figure}
\inputtikz{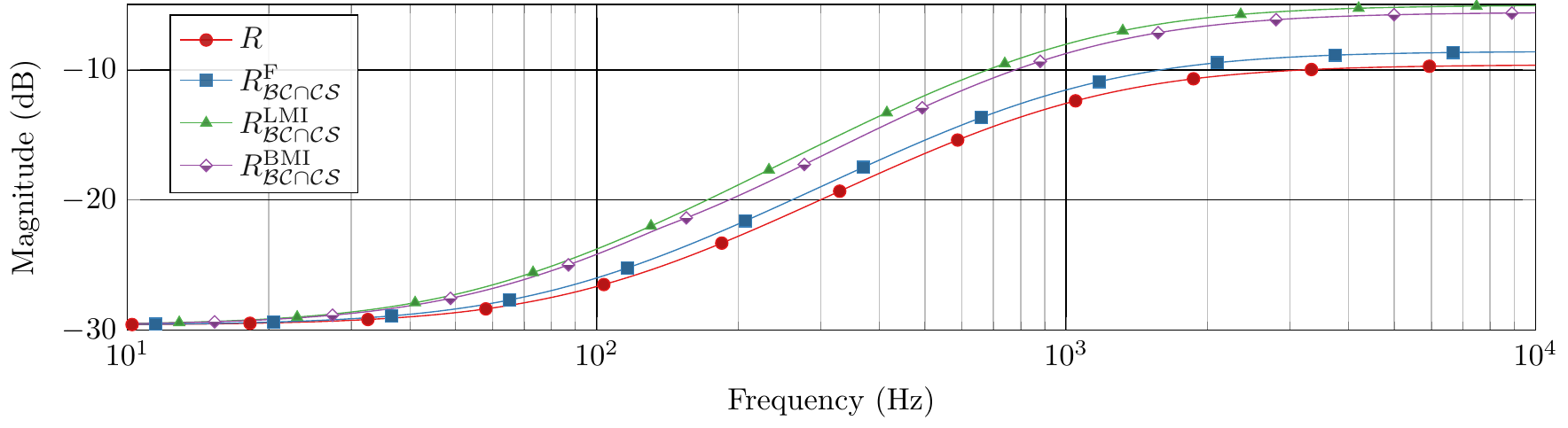}%
\caption{Upper bound on unkown uncertainty $\twonorm{\Theta(\jw)}$ for the original system and the $\BC\cap\CS$ approximations of the ALBA synchrotron.\label{fig:uncertainty}}%
\end{figure}

The robustness of the system is measured by the frequency-dependent upper bound~\eqref{eq:normcondrob} on the unknown additional uncertainty $\Theta(\jw)$, which is shown in Fig.~\ref{fig:uncertainty} for the system that uses the original $R$ (\addlegendimageintext{mark=*,draw=legreddraw,fill=legredfill}) and the different approximations from Section~\ref{sec:approx:application:frob}. 

At low frequencies ($\omega\leq 2\pi\times\SI{20}{\radian\per\second}$), the norm of the admissible unknown uncertainty is at least $\SI{-30}{\dB}\approx 0.03$ before the closed-loop system might become unstable, and this upper bound is of similar magnitude for all systems from Fig.~\ref{fig:uncertainty}, including the one that uses the original $R$. For $\omega\rightarrow 0$, the right-hand side of the upper bound~\eqref{eq:normcondrob} converges to $1/\twonorm{R^\dagger}$ for $Q(s)$ that uses $R$ and $1/\twonorm{(R_\mathcal{S}^{(\cdot)})^\dagger\inv{(I+\Phi_\mathcal{S}^{(\cdot)})}}$ for $Q(s)$ that uses $R_\mathcal{S}^{(\cdot)}$. 

At high frequencies ($\omega\geq\SI{1}{\kHz}$), the approximations $R_{\BC\cap\CS}^\text{LMI}$ (\addlegendimageintext{mark=triangle*,draw=leggreendraw,fill=leggreenfill}) and $R_{\BC\cap\CS}^\text{BMI}$ (\addlegendimageintext{mark=halfdiamond*,draw=legvioletdraw,fill=legvioletfill}) yield significantly more robust systems than the systems that use $R$ and $R_{\BC\cap\CS}^\text{F}$. At \SI{1}{\kHz}, the admissible uncertainty is at least $\SI{6}{\dB}\approx 2$ for the $R_{\BC\cap\CS}^\text{LMI}$ and $R_{\BC\cap\CS}^\text{BMI}$ approximations, which suggests that the performance loss from Fig.~\ref{fig:sensitivity} is traded against the gain in robustness from Fig.~\ref{fig:uncertainty}.
%
\section{Conclusion\label{sec:approx:conclusion}}
In this paper, an IMC structure was used to control CD systems with approximate structural symmetries. After fixing the controller structure and substituting a generic approximation for the original plant model, the properties of the resulting closed-loop were analysed. Based on this analysis, conditions on stability, performance and robustness were derived that can be embedded in an SDP with the aim of finding an approximation that has exact structural symmetries.

In contrast to SDP-based approximations, the Frobenius norm approximation benefits from a closed-form solution and a block-hollow structure of the resulting approximation error. Based on the properties of the approximation error, a simple block-diagonal dominance condition was derived to verify whether a CD system is amenable to a Frobenius norm approximation. In general, row or column block-diagonal dominance of the steady-state gain matrix suffices for closed-loop stability of the symmetric approximation.

It was also shown that the Frobenius norm approximation can be suboptimal in the sense that it can yield an unstable system or a system with poor performance. In this case, the SDP-based approach can be used to find approximations with improved performance and robustness properties. For the case that the Frobenius approximation yields an unstable closed loop, it was shown how to define an optimisation problem to find a stabilising approximation (if it exists). If the Frobenius norm approximation yields a stable closed loop, it can be used to initialise the SDP-based problems that can lead to approximations with better robustness and performance properties. These optimisation problems can be solved in the symmetric domain where the matrices are sparse, which allows for large-scale systems with large optimisation problems to be investigated that would otherwise be difficult to solve if all matrices were dense.

The asymmetry of the steady-state gain matrix of a CD system has been investigated, but a possible asymmetry of the actuator dynamics has been ignored. For systems with asymmetric actuator dynamics, the nominal stability condition, which is based on evaluating the spectral radius of a static closed-loop matrix, would need to be evaluated on a frequency-by-frequency basis. It is unclear whether the block-diagonal dominance condition for stability remains sufficient. Future research could extend the framework to allow for asymmetry in the actuator dynamics.

Certain cross-directional systems, such as synchrotron light sources, suffer from an ill-conditioned steady-state gain matrix. In this case, the controller produces large actuator gains in direction of small-magnitude singular values and the control system becomes sensitive to modelling errors. In practice, a static regularisation matrix is added to the IMC structure, which damps the controller gains in direction of the small-magnitude singular values. The regularisation gain has been omitted from this analysis, which could be considered in future research directions.
\begin{ack}
The authors would like to thank U. Iriso of ALBA for providing the orbit response matrix. This work was supported in part by the Diamond Light Source and in part by the Engineering and Physical Sciences Research Council (EPSRC) with a Collaborative Awards in Science and Engineering (CASE) studentship.
\end{ack}
\appendix
\section*{Appendices}
\section{Bilinear Matrix Inequalities\label{app:bmi}}
\subsection{Convexifying~\eqref{eq:NS2}}
The BMI~\eqref{eq:NS2} is given as
\begin{align*}
\mathcal{F}(X)=-
\begin{bmatrix}
I & (RX-I)^2\\
\hermbr{\dots} & I
\end{bmatrix}\preceq 0.
\end{align*}
The matrix functional $\mathcal{F}(X)$ is concave iff \[\mathcal{F}((1-\alpha) X + \alpha Y)\succeq (1-\alpha)\mathcal{F}(X)+\alpha\mathcal{F}(Y),\] for $\alpha\in[0,1]$. Here, the concavity condition is evaluated as
\begin{align*}
&\mathcal{F}((1-\alpha) X + \alpha Y) - (1-\alpha)\mathcal{F}(X)-\alpha\mathcal{F}(Y)
=(\alpha-\alpha^2)\begin{bmatrix}
0 & (RX+RY-2I)^2\\
\hermbr{\dots} & 0
\end{bmatrix},
\end{align*}
which shows that $\mathcal{F}(X)$ is \emph{not} a concave function. For~\eqref{eq:NS2}, Alg.~\ref{alg:convexifying} therefore only generates feasible iterates, without necessarily converging to a local optimum.

The convexifying potential matrix functional $G(X,X_k)$ can be chosen as
\begin{align*}
G(X,X_k)\eqdef
&\begin{bmatrix}I \\ 0\end{bmatrix}
(RX-RX_k)^2
\begin{bmatrix}0 & I\end{bmatrix} +
\begin{bmatrix}0 \\ I\end{bmatrix}
\hermbr{(RX-RX_k)^2}
\begin{bmatrix}I & 0\end{bmatrix}.
\end{align*}
Note that $G(X,X_k)$ is a quadratic form in $X$ and $X_k$ and therefore satisfies the assumptions from Def.~\ref{def:convexmatfunc}. The sum $\mathcal{F}(X)+G(X,X_k)$ is obtained as
\begin{align*}
\mathcal{F}(X)+G(X,X_k)=\begin{bmatrix}
-I & R(2X\!+\!X_kRX_k\!-\!XRX_k\!-\!X_kRX)\!-\!I\\
\hermbr{\dots} & -I
\end{bmatrix},
\end{align*}
which results in an LMI in $X$.
\subsection{Convexifying~\eqref{eq:NS3}}
The matrix inequality~\eqref{eq:NS3} is given as
\begin{equation*}
\mathcal{F}(X,P)=-
\begin{bmatrix}
\inv{P} & RX-I\\
\hermbr{RX-I} & P
\end{bmatrix}\prec 0,
\end{equation*}
which, after pre- and post-multplying with $\diag(P,I)$ can be interpretated as a BMI in $X$ and $P$. In~\cite{CONVEXBMI}, it is shown that $\mathcal{F}(X,P)$ is a concave matrix functional and that with
\begin{align*}
&G(P,P_k)\eqdef\begin{bmatrix} I\\ 0\end{bmatrix}\hermbr{\inv{P}_k- \inv{P}}P\br{\inv{P}_k- \inv{P}}\begin{bmatrix} I & 0\end{bmatrix},
\end{align*}
one obtains the sum $\mathcal{F}(X,P)+G(P,P_k)$ as
\begin{align*}
\mathcal{F}(X,P)+G(P,P_k)=\begin{bmatrix}\inv{P}_k ( P-2 P_k)\inv{P}_k & -(RX-I)\\ -\hermbr{RX-I} & -P\end{bmatrix}\prec 0,
\end{align*}
which, to avoid numerical difficulties with computing $\inv{P}_k$, can be reformulated as
\begin{align*}
\inv{P}_k=\begin{bmatrix}( P-2 P_k) & -P_k(RX-I)\\ -\hermbr{RX-I}P_k & -P\end{bmatrix}\prec 0.
\end{align*}
\subsection{Convexifying~\eqref{eq:NP1}}
To convexify~\eqref{eq:NP1}, the Schur complement is applied to $\twonorm{T(\jw){\Phi}_\mathcal{S}^2-{\Phi}_\mathcal{S}}^2\leq\alpha_\omega^2$. The matrix inequality~\eqref{eq:NP1} is then reformulated as
\begin{align*}
\mathcal{F}(X) = \begin{bmatrix}-I & \mathcal{R}-T(\jw)\mathcal{R}^2\\ \hermbr{\dots} & -\alpha_\omega^2 I\end{bmatrix}\preceq 0,
\end{align*}
where $\mathcal{R}\eqdef RX-I$. A matrix functional $G(X,X_k)$ that convexifies the BMI is given by
\begin{align*}
G(X,X_k)\eqdef
&\begin{bmatrix}I \\ 0\end{bmatrix}
T(\jw)(RX-RX_k)^2
\begin{bmatrix}0 & I\end{bmatrix} +
\begin{bmatrix}0 \\ I\end{bmatrix}
\hermbr{T(\jw)(RX-RX_k)^2}
\begin{bmatrix}I & 0\end{bmatrix},
\end{align*}
and the sum $\mathcal{F}(X)+G(X,X_k)$ evaluates to
\begin{equation*}
\begin{aligned}
\mathcal{F}(X)+G(X,X_k)=\begin{bmatrix}
-I & \mathcal{R}+T(\jw)(\mathcal{R}_k^2-\mathcal{R}_k\mathcal{R}-\mathcal{R}\mathcal{R}_k)\\
\hermbr{\dots} & -\alpha_\omega^2 I
\end{bmatrix}\preceq 0,
\end{aligned}
\end{equation*}
where $\mathcal{R}_k\eqdef RX_k-I$.
\balance
\bibliographystyle{apalike-ampersand}
{\scriptsize
\bibliography{master_bib_abbrev}}
\end{document}